# First-principles calculation of defect free energies:

## General aspects illustrated in the case of bcc-Fe

D. Murali, M. Posselt[*], and M. Schiwarth

Helmholtz-Zentrum Dresden - Rossendorf,

Institute of Ion Beam Physics and Materials Research,

Bautzner Landstraße 400, 01328 Dresden, Germany

**Abstract**

Modeling of nanostructure evolution in solids requires comprehensive data on the properties of defects such as the vacancy and foreign atoms. Since most processes occur at elevated temperatures not only the energetics of defects in the ground state but also their temperature-dependent free energies must be known. The first-principles calculation of contributions of phonon and electron excitations to free formation, binding, and migration energies of defects is illustrated in the case of bcc-Fe. First of all, the ground state properties of the vacancy, the foreign atoms Cu, Y, Ti, Cr, Mn, Ni, V, Mo, Si, Al, Co, O, and the O-vacancy pair are determined under constant volume (CV) as well as zero pressure (ZP) conditions, and relations between the results of both kinds of calculations are discussed. Second, the phonon contribution to defect free energies is calculated within the harmonic approximation using the equilibrium atomic positions determined in the ground state under CV and ZP conditions. In most cases the ZP-based free formation energy decreases monotonously with temperature, whereas for CV-based data both an increase and a decrease were found. The application of a quasi-harmonic correction to the ZP-based data does not modify this picture significantly. However, the corrected data are valid under zero-pressure conditions at higher temperatures than in the framework of the purely harmonic approach. The difference between CV- and ZP-based data is mainly due to the volume change of the supercell since the relative arrangement of atoms in the environment of the defects is nearly identical in the two cases. A simple transformation similar to the quasi-harmonic approach is found between the CV- and ZP-based frequencies. Therefore, it is not necessary to calculate these quantities and the corresponding defect free energies separately. In contrast to ground state energetics the CV- and ZP-based defect free energies do not become equal with increasing supercell size. Third, it was found that the contribution of electron excitations to the defect free energy can lead to an additional deviation of the total free energy from the ground state value or can compensate the deviation caused by the phonon contribution. Finally, self-diffusion via the vacancy mechanism is investigated. The ratio of the respective CV- and ZP-based results for the vacancy diffusivity is nearly equal to the reciprocal of that for the equilibrium concentration. This behavior leads to almost identical CV- and ZP-based values for the self-diffusion coefficient. Obviously, this agreement is accidental. The consideration of the temperature dependence of the magnetization yields self-diffusion data in very good agreement with experiments.

[*] Corresponding author,

Address: Helmholtz-Zentrum Dresden - Rossendorf, Bautzner Landstraße 400,

01328 Dresden, Germany

Electronic address: M.Posselt@hzdr.de,

Phone: +49 351 260 3279,

Fax: +49 351 260 3285

# I. INTRODUCTION

Processes at elevated temperatures can lead to nanostructure evolution and may cause the modification of mechanical, electrical and magnetic properties of solids. Together with vacancies and self-interstitials foreign atoms play an important role in the alteration of the nanostructure. Experimental investigations on defect or nanostructure thermodynamics and kinetics require a sufficiently high temperature to observe the process of interest in a finite time. On the other hand, first-principle calculations using Density Functional Theory (DFT) have been widely used to determine the *ground state* properties of point defects, foreign atoms, and other nanoobjects. The data obtained for formation, migration and binding energies are often used in multiscale modeling of processes at *elevated temperatures*, by applying kinetic Monte Carlo simulations (cf. e.g. Refs. 1-4) or rate theory. This is clearly not a fully consistent approach. It is only recently that authors have pointed out the importance of considering full DFT-based free energy data at nonzero temperatures instead of those obtained from the energetics in the ground state.[5-21] Only few studies on this topic were performed earlier.[22-27] In general excitations of electrons, phonons, magnons and other quasi-particles can contribute to the free energy of defects at elevated temperatures.[7,10,11,13] In the case of an extended spatial configuration of a defect the contribution of its "internal" configurational entropy must be considered as well in these calculations. An example is the dumbbell self-interstitial in bcc-Fe which exists in six equivalent orientations.

In one of the first full DFT-based study on the free energy of foreign atoms in bcc-Fe Reith *et al.*[6] showed that the observed thermodynamic stability of Cu substitutional atoms cannot be properly explained using the DFT formation energy in the ground state, whereas a better agreement was obtained if the contribution of phonon excitations to the free formation energy is taken into account. The effect of the vibrational contribution to the nucleation free energy of Cu clusters in bcc-Fe was clearly evidenced in the work of Yuge *et al.*[26] DFT calculations of Lucas *et al.*[5] demonstrated that even at moderate temperatures the phonon contribution to the free formation energy of vacancies and self-interstitials in bcc-Fe must not be neglected. Huang *et al.*[7] performed DFT studies on self-diffusion in bcc-Fe and showed that the consideration of phonon and electron contributions to the free formation and migration energy of the vacancy led to a very good agreement with experimental data. Using this methodology they could also reproduce the experimental diffusion coefficients of W and Mo in bcc-Fe. A similar procedure to determine free energies was also used by Mantina *et al.*[8] and Tucker *et al.*[10] to calculate self- and impurity diffusion in Al and Ni, respectively. Satta *et al.*[23,24] investigated the contribution of electronic excitations to the activation energy of self-diffusion in W and Ta and found that in the former case the effect is more pronounced than in the latter. The important role of vibrational contributions to the free formation energy and entropy of defects was also shown in DFT calculations for non-metallic materials, namely, for the O vacancy and the Zn interstitial in ZnO (Ref. 14), for the O vacancy in Sr-doped complex perovskites[15], as well as for native point defects in $Cu_2ZnSnS_4$ (Ref. 16), $In_2O_3$ (Ref. 20), and SiC (Ref. 21). Furthermore, several authors demonstrated that the effect of phonon excitations must be considered in first-principles calculations of phase diagrams of binary and ternary metallic and non-metallic alloys.[17-19,22,27] While in most of the above-mentioned studies vibrational effects were treated within the harmonic or quasi-harmonic approximation anharmonic effects have been considered in a very recent paper.[11] These authors have investigated the temperature dependence of the free formation energy of the vacancy in Al and Cu and have found that the formation entropy is not constant, as often assumed, but increases with temperature. This leads to a nonlinear temperature dependence of the free formation energy. The findings of Ref. 11 have contributed to the explanation of apparent discrepancies between DFT results and experimental data.

The focus of the present paper is on the methodology of the calculation of defect free energies within the framework of the supercell approach *and* on the application of this procedure to bcc-Fe. After introduction of a foreign atom or an intrinsic defect into the supercell two alternative kinds of calculations may be performed to determine the energetics in the ground state: (i) The positions of atoms are relaxed but the volume and shape of the supercell are held constant (constant volume calculations - CV). (ii) Both the positions of atoms and the volume and shape of the supercell are relaxed so that the total pressure or stress on the supercell is zero (zero pressure calculations - ZP). In the present work CV as well as ZP is performed and the relation between the results obtained by the two methods is discussed, with a focus on the dependence on the supercell size. Despite the wealth of literature on DFT calculations on ground state energetics of defects such a discussion was performed only in few previous papers. The positions of atoms in the supercell obtained by CV and ZP are used to determine the contribution of phonon excitations to free energies by relatively time-consuming, state-of-the-art calculations. A detailed discussion on the relation between the CV- and ZP-based defect free energies and their size dependence is performed because this was missing in the few previous papers on corresponding DFT calculations. In selected cases a quasi-harmonic correction is applied to ZP-based defect free energies. Since experiments are usually performed at zero external pressure these results are most interesting for a comparison. The electronic contributions to the free energy are treated separately from those of phonons within the meaning of the adiabatic approximation.[28] Contributions due to magnon excitations are not considered.

In the present work the ground state energetics as well as the vibrational and electronic contributions to the free formation energy of the vacancy (v), of the foreign atoms Cu, Y, Ti, Cr, Mn, Ni, V, Mo, Si, Al, Co, O, and of the O-v pair in bcc-Fe are determined. Furthermore, the free binding energy of O-v, the free migration energy of the vacancy, and the self-diffusion coefficient in Fe are calculated. The results of present investigations shall contribute to a better understanding and an improved modeling of nanostructure evolution in ferritic Fe and Fe-Cr alloys which are important for various practical applications, in particular as basic structural materials for present and future nuclear fission and fusion reactors. For example, Cu, Ni, and Mn are characteristic solutes in conventional reactor pressure vessel steel, and Cr, Y, O, and Ti are basic alloying elements in oxide dispersion strengthened ferritic Fe-Cr steel that shows an excellent high-temperature creep strength and a high radiation resistance.[29] The solutes Mo, Al, Si, V, and Co can be present in both ferritic Fe and Fe-Cr alloys. The vacancy plays a fundamental role in the diffusion of most of the foreign atoms and in the formation of nanoclusters containing these species. It should be noticed that in all these applications the relevant temperatures are between about 600 and 1000 K.

The paper is organized as follows. The ground state energetics of defects is considered in Sec. II. This includes the calculation of the formation energy of the vacancy, foreign atoms and of the O-v pair, the vacancy migration energy as well as the binding energy of O-v. Secs. III and IV deal with the effect of excitations of phonons and electrons on the free formation, migration and binding energy. Both sections can be considered as the main part of the work. The free formation and migration energies of the vacancy determined by DFT methods in the preceding sections are used in Sec. V to obtain the self-diffusion coefficient in bcc-Fe. The calculated data are compared with results of previous DFT calculations and available experimental data.

## II. GROUND STATE ENERGETICS

### A. Calculation method

The Vienna ab-initio simulation package (VASP)[30,31] was used to perform DFT calculations. Plane wave basis sets and pseudopotentials generated within the projected augmented wave (PAW) approach were employed. For transition metals and magnetic systems the PAW method is generally preferred compared to the ultra-soft pseudopotential (USPP) approach.[32-34] In all calculations the spin polarized formalism was applied and the generalized gradient approximation GGA-PBE[35] was used to describe the exchange and correlation effects. The Brillouin zone sampling was performed using the Monkhorst-Pack scheme.[36] In all cases a plane wave cutoff of 500 eV was used. Calculations on the energetics of foreign atoms, the vacancy (v), and the oxygen-vacancy pair (O-v) were performed for a bcc-Fe supercell with 54 lattice sites and Brillouin zone sampling of $6\times6\times6$ $k$ points as well as for a cell with 128 sites and $3\times3\times3$ $k$ points. In selected cases supercells with 250 sites and $2\times2\times2$ $k$ points were considered. After introduction of the defect into the supercell two types of calculations were carried out in order to determine the energetics in the ground state: (i) The positions of atoms were relaxed at constant volume and shape of the supercell (constant volume calculations - CV). (ii) After step (i) the positions of atoms as well as the volume and shape of the supercell were relaxed so that the total stress/pressure on the supercell became zero (zero pressure calculations - ZP). Special care was taken to ensure a high precision of the results. This was achieved by performing several calculations to obtain the most suitable parameters for the iterations carried out in VASP. There are two important criteria to be considered: (i) First, CV and ZP are carried out until the residual force acting on any atom falls below a given threshold, and (ii) at each step of CV and ZP the relaxation of the electronic degrees of freedom is performed until the total energy change falls below another threshold. It was found that the optimum threshold values depend on the example considered. They vary between $10^{-2}$ and $10^{-5}$ eV/Å in the first case and between $10^{-5}$ and $10^{-7}$ eV in the second case.

In the ground state the formation energy of a substitutional defect $X$ is defined by

$$E_f(X) = E_D(X) - E_0 + \mu_0 - \mu_X \qquad (1)$$

(cf. Refs. 13, and 37-39) where $E_D(X)$ and $E_0$ denote the total energy of the supercell with and without the defect, respectively. The quantities $\mu_0$ and $\mu_X$ are the chemical potentials of the Fe and the $X$ atoms, respectively, in the corresponding reservoirs or reference systems. In the case of a very dilute alloy $\mu_0$ can be set equal to the energy per atom $E_{0,A}$ in perfect bcc-Fe. According to the common practice in DFT calculations of defect energetics in bcc-Fe [2,5,6,7,32,34,40-49] in the present work the perfect crystal of the element $X$ (existing under standard conditions) is chosen as reservoir or reference system. Then, the formation energy of a foreign atom $X$ on a lattice site is calculated by

$$E_f(X) = E_D(X) - (N-1)E_{0,A} - E_{X,A} \tag{2}$$

where $N$ and $E_{X,A}$ are the number of lattice sites in the supercell and the energy per atom in in the bulk reference crystal of the element $X$, respectively. In the case of the vacancy the formation energy is defined by a similar relation but without the term $E_{X,A}$. In the present work the following reference systems are used to determine the formation energy of the corresponding foreign atom in bcc-Fe: (i) hcp-Y, (ii) hcp-Ti, (iii) bcc-Cr, (iv) fcc-Cu, (v) Mn (cubic), (vi) fcc-Ni, (vii) bcc-V, (viii) bcc-Mo, (ix) Si (diamond structure), (x) fcc-Al, (xi) bcc-Co, and (xii) FeO (NaCl structure). The ground states determined for Ni and Co are ferromagnetic while the ground states of Cr, and FeO are antiferromagnetic. The other reference systems have nonmagnetic ground states. Table I shows the lattice parameters of bcc-Fe and the reference materials obtained by present DFT calculations in comparison with experimental data. A good agreement is found. The number of atoms in the supercell and the number of $k$ points used in the calculation are also given in Table I. A relatively large size of the supercell is required to obtain a high accuracy of the phonon frequencies (cf. Sec. III). The table contains as well the notations of the PAW-PBE pseudopotentials employed in this work.

The formation energy of an O interstitial atom in bcc-Fe is determined using the relation

$$E_f(O) = E_D(O) - (N-1)E_{Fe,A} - E_{FeO,FU} \quad (3)$$

where $E_{FeO,FU}$ is the energy of FeO per formula unit. The formation and binding energy of the O-v pair are calculated using

$$E_f(\text{O-v}) = E_D(\text{O-v}) - (N-2)E_{Fe,A} - E_{FeO,FU} \quad (4)$$

$$E_{bind}(\text{O-v}) = E_f(\text{O-v}) - E_f(O) - E_f(v) \quad (5)$$

By definition the value of $E_{bind}$ is negative if attraction between O and v dominates.

The energy barrier for vacancy migration in bcc-Fe was evaluated using the Nudged Elastic Band method[62,63] by considering the exchange of a Fe atom with the nearest neighbor vacancy. The states with the initial and final positions of the vacancy were connected by a number of system images that were constructed along the reaction path. These images were determined by (restricted) energy minimization under both CV and ZP conditions. In order to find an exact saddle point with a minimum number of images the climbing image method was employed as implemented in the "vtsttools".[64]

### B. Formation energy of the vacancy and foreign atoms

Table II shows the calculated formation energies $E_f$ of the vacancy, Cu, Y, Ti, Cr, Mn, Ni, V, Mo, Si, Al, Co, O, and the O-v pair in bcc-Fe. The most favorable position of the O atom is the octahedral interstitial site while the other species prefer regular lattice sites. For all these examples the data obtained by CV and ZP are presented for supercells with 54 and 128 bcc sites. The formation energy of the vacancy and the Cu atom was also determined for the supercell with 250 sites. A clear quantitative comparison of the $E_f$ data obtained for the different defects is not simple since this depends on the choice of the reference system. In the present work the reference systems were chosen in the same or similar manner as in related papers dealing with such defects Fe.[2,5,6,7,32,34,40-49] In most cases the reference crystal was selected to model the condition of a very dilute Fe alloy where the solubility or equilibrium concentration of the foreign atom $C_{eq}(X)$ is directly related to the total free formation energy $F_f^{tot}(X,T)$ (cf. e.g. Refs. 13, 38, and 39)

$$C_{eq} = \exp\left(-\frac{F_f^{tot}(X,T)}{k_B T}\right) \quad (6)$$

with

$$F_f^{tot}(X,T) = E_f(X) + F_f^{vib}(X,T) + F_f^{el}(X,T) \quad (7)$$

$F_f^{vib}(X,T)$ and $F_f^{el}(X,T)$ denote contributions of phonon and electron excitations to $F_f^{tot}(X,T)$ (cf. Secs. III and IV). If one ignores the quantities $F_f^{vib}(X,T)$ and $F_f^{el}(X,T)$ the equilibrium concentration is approximately determined by the value of $E_f(X)$. The comparison with available phase diagrams[65] shows that formation energy data given in Table II follow a similar trend as the solubility of the foreign atoms in bcc-Fe. A low formation energy correspond to a much higher solubility than a high formation energy. Amongst the foreign atoms considered in Table II, Y and O have the highest formation energies. According to the phase diagrams these atoms have also the lowest solubility in bcc-Fe. Regarding the fact that that in Table II certain values of $E_f(X)$ are negative we refer to the corresponding results and their discussion in the literature.[32,40,41,44,45,47,49,66,67]

In Table II DFT data from the literature are shown as well. Most of them were obtained by CV. Apart from Cr a good agreement with results of the present work is obtained if PAW pseudopotentials were employed in these calculations [5-7,32,34,40-42,46,49,68] while a larger difference is found if the less accurate USPP pseudopotentials were used.[43-45] Similar differences were also found by other authors.[6,33,34] The difference to Olsson's PAW-based value for the substitutional energy of Cr (Refs. 32 and 68) could be explained by the fact that in the present work another pseudopotential was used for Cr (cf. Table I and Ref. 68).

In many cases results of this work obtained for supercells with 54, 128, and 250 bcc sites are nearly equal, both for CV and ZP. This is an indication that the choice of a supercell with 54 bcc sites already leads to reasonable results. Moreover, for a given supercell size results of CV and ZP are often very similar. The total internal pressure in the supercell obtained in the case of CV is given in Table III. If the ZP procedure is applied the volume of the supercell expands or contracts depending on the type of the point defect. The corresponding data are also given in Table III. While the introduction of v and Si leads to a contraction the other foreign atoms cause an expansion. As expected, the internal pressure $p$ and the relative change of the cell volume $(V-V_0)/V_0$ are considerably reduced if the supercell size increases. On the other hand the absolute volume difference $(V-V_0)$ does not change significantly with supercell size, which is another indication that a cell with 54 lattice sites is often an acceptable choice. In most cases an isotropic expansion or contraction is observed so that the cubic shape of the supercell is conserved. For the O octahedral interstitial a larger expansion in the direction of the two nearest neighbor Fe atoms than in the other two directions is obtained by ZP. Such a tetragonal distortion was also observed for other foreign atoms on octahedral sites, in particular for carbon.[69] In general it is found that the additional relaxation of volume and shape of the supercell leads to a slight decrease of the formation energy. In the case of Y and O a supercell with 128 bcc sites is obviously not large enough to yield completely converged data for the formation energy. The atomic size of Y is very large compared to the Fe host atoms and O is an "extra" atom in the bcc lattice on the octahedral interstitial position. Therefore, for Y and O the total internal pressure observed for CV and the expansion of the supercell found by ZP are highest compared to the other foreign atoms (cf.

Table III). The difference between the formation energy obtained by CV and ZP can be determined by[70]

$$E_f^{CV} - E_f^{ZP} = \frac{V}{2} C_{ijkl}\, \varepsilon_{ij}\, \varepsilon_{kl} \tag{8}$$

where $\varepsilon_{mn}$ is the homogeneous strain on the supercell in the case of ZP and $C_{ijkl}$ is the tensor of the elastic constants. If ZP leads to an isotropic expansion or contraction this difference is given by

$$E_f^{CV} - E_f^{ZP} = \frac{1}{2}\left(\frac{V-V_0}{V_0}\right)^2 B_0 V = \frac{1}{2} p^2 \frac{V}{B_0} = \frac{1}{2} p\,(V-V_0) \frac{V}{V_0} \tag{9}$$

where $B_0$ is the bulk modulus of bcc-Fe. In the case of a tetragonal distortion of the supercell a slightly different relation is valid. Taking the values of $E_f^{CV}$ and $E_f^{ZP}$ from Table II and the pressure and volume data from Table III this kind of relation was employed to confirm that the results obtained by CV and ZP are fully consistent with each other. Details can be found in the Supplemental Material.[71]

Eq. (9) clearly illustrates that $E_f^{CV} - E_f^{ZP}$ tends towards zero with increasing supercell size, $E_f^{CV} - E_f^{ZP} \propto 1/N$, since $(V-V_0/V_0) \propto 1/N$ and $V \propto N$, $V_0 \propto N$. That means that results of CV and ZP become equal for a sufficiently large supercell, i.e. in the case of a macroscopic crystal (cf. also Ref. 70).

### C. Energetics of the O-v pair

The formation energy of the O-v pair at a distance equal to about a half of the lattice parameter is also given in Table II. The difference between the values of $E_f(\text{O-v})$ determined by CV and ZP is smaller than for $E_f(\text{O})$. The binding energy of the O-v pair determined by Eq. (5) is -1.74 and -1.53 eV if formation energies calculated by CV and ZP are used, respectively. The latter number agrees well with the value of -1.44 eV obtained by Barouh *et al.*[72] using the SIESTA code and zero pressure conditions while the former is in the range of the literature data obtained under constant volume conditions (Ref. 42: -1.45 eV, Ref. 40: -1.55 eV, Ref. 49: -1.65 eV, Ref. 41: -1.7 eV). Obviously, the O-v pair is energetically highly favored compared to the coexistence of a single vacancy and a single O interstitial. More distant O-v pairs show a significantly lower binding. These results are fully consistent with findings of the other researchers.[40-42,49,72] The energy gain by the pair formation is due to the reconfiguration of the electronic structure and the reduction of the elastic strain. The latter can be clearly seen by comparing the corresponding lines in Table III. Similarly to the case of the O interstitial a tetragonal distortion of the supercell is found by ZP.

### D. Migration energy of the vacancy

The energy barrier $E_{mig}$ for vacancy migration in bcc Fe was found to be 0.68 eV both for CV and ZP, for a supercell containing 54 bcc sites. This is similar to the case of vacancy formation energy where the difference between results of CV and ZP is very small. The above value for the migration energy agrees well with data obtained by previous DFT calculations (Ref. 7: 0.64 eV, Ref. 43: 0.65 eV, Ref. 46: 0.67 eV, Ref. 47: 0.68 eV, Ref. 48: 0.66 eV, Ref. 73: 0.67 eV).

## III. EFFECT OF PHONON EXCITATIONS

### A. Computational procedure

In order to determine the contribution of phonon excitations to the free energy, the vibrational frequencies of the corresponding supercells were calculated using the method implemented in the VASP code. This procedure employs the harmonic approximation and the frozen phonon approach (cf. Refs. 7, 74, and 75). In the present work two kinds of calculations were performed for the supercells with defects: The dynamical matrix with the force derivatives was determined using ground-state atomic positions obtained either by the CV or ZP. In the VASP code the calculation of phonon frequencies is done only for the gamma point in the space of the phonon wave vectors. Therefore, a reasonably large supercell size must be chosen both for the cell with the defect and with the reference material (cf. Table I). Most of the vibrational calculations for supercells with a defect were performed using 54 bcc sites and a Brillouin zone sampling of $6 \times 6 \times 6$ $k$ points. In order to check the convergence of the vibrational contribution to the free energy with respect to supercell size, some calculations were also carried out for supercells with 128 sites and $3 \times 3 \times 3$ $k$ points as well as with 250 sites and $2 \times 2 \times 2$ $k$ points.

The diagonalization of the dynamical matrix yields all vibrational frequencies $\omega_i$ of the supercell. Within the harmonic approximation the vibrational free energy $F^{vib}(T)$ of a system of $N$ atoms is given by

$$F^{vib}(T) = \sum_{i=1}^{3N-3} \left[ \frac{1}{2}\hbar\omega_i + k_B T \ln\left(1 - e^{-\frac{\hbar\omega_i}{k_B T}}\right) \right] \tag{10}$$

$$F^{vib}(T) = U^{vib}(T) - T S^{vib}(T) \tag{11}$$

where $U^{vib}(T)$ and $S^{vib}(T)$ denote the vibrational contribution to internal energy and entropy, respectively, and $k_B$ is the Boltzmann constant. $S^{vib}(T)$ can be obtained by

$$S^{vib}(T) = -\frac{\partial F^{vib}}{\partial T} \tag{12}$$

The vibrational contribution to the free formation energy of a foreign atom $X$ on a lattice site is determined similarly to Eq. (2)

$$F_f^{vib}(T) = F_D^{vib}(T) - (N-1)F_{0,A}^{vib}(T) - F_{X,A}^{vib}(T) \tag{13}$$

(cf. Refs. 5 and 6) where $F_D^{vib}(T)$ denotes the vibrational free energy of the bcc-Fe supercell with the foreign atom $X$ and $N$ lattice sites, whereas $F_{0,A}^{\text{vib}}(T)$ and $F_{X,A}^{\text{vib}}(T)$ are the vibrational free energy per atom in perfect bcc-Fe and in the bulk reference crystal of the element $X$, respectively. Modified relations must be used for the vacancy and the O atom. The vibrational contribution to the free formation and binding energy of the O-v pair is calculated in a manner similar to Eqs. (4,5). The vibrational contribution to the formation entropy is determined by a relation similar to Eq. (12).

The vibrational contribution to the free migration energy of the vacancy is calculated by the difference between the vibrational free energy at the saddle point and at the equilibrium state of the vacancy.

$$\Delta F_{\text{mig}}^{\text{vib}}(T) = F_{SP}^{\text{vib}}(T) - F_{0}^{\text{vib}}(T) \tag{14}$$

The quantities on the right-hand side of Eq. (14) are determined using Eq. (10) with the exception that in the first term the sum is over $3N-4$ vibrational degrees of freedom. In calculations of $\Delta F_{\text{mig}}^{\text{vib}}(T)$ atomic positions determined by both CV and ZP were used. Within the harmonic transition state theory (cf. Refs. 10, 76, and 77) the vacancy diffusivity in the bcc lattice is obtained by

$$D(T) = a^2 \frac{k_B T}{h} \exp\left(-\frac{\Delta F_{\text{mig}}^{\text{vib}}(T)}{k_B T}\right) \exp\left(-\frac{E_{\text{mig}}}{k_B T}\right) \tag{15}$$

where $h$ is Planck's constant.

At this point the further use of the acronyms CV and ZP must be explained. In this work the contributions of phonon and electron excitations to defect free energies are determined for the two different systems (supercells with defects) obtained by energy minimization in the ground state under CV and ZP conditions. The corresponding results are therefore also denoted by the terms CV and ZP. However, the ZP-based data for the vibrational frequencies and the corresponding contributions to the free energy can only describe approximately the zero pressure case, namely at low temperatures. Additionally, the ZP-based vibrational frequencies are corrected according to the quasi-harmonic approach (cf. e.g. Ref. 78 and 79)

$$\omega_i^{ZP,qh} = \omega_i^{ZP}\left(1 - \gamma \frac{\Delta V}{V}\right) \qquad (16)$$

This relation allows the determination of frequencies for zero-pressure conditions at higher temperatures than the purely harmonic approach but it is limited to temperatures, where anharmonic effects do not prevail. The calculation of the volume increase $\Delta V$ due to thermal expansion, and also due to the small effect of zero point vibrations, is explained in the Supplemental Material.[80] The quantity $\omega_i^{ZP}$ denotes ZP-based frequencies determined within the harmonic approximation, i.e. using the dynamical matrix, $\gamma$ is the thermodynamic or phonon Grüneisen parameter of bcc-Fe, and $V$ is the volume of the supercell with the defect determined by ZP in the ground state (cf. Table III). In general Eq. (16) holds if $\Delta V/V$ is sufficiently small. In the following the quasi-harmonic approach (16) is applied to correct the ZP-based vibrational contributions to defect free energies in selected cases. In order to obtain these data, because of Eq. (13) the quasi-harmonic approach must be also employed in the calculation of the frequencies and the vibrational free energy of the supercell containing the reference material. Since the majority of experiments are performed at zero external pressure and elevated temperatures the defect free energy data resulting from the quasi-harmonic correction are most interesting for a comparison. In this work the quasi-harmonic approach [Eq. (16)] is used under the assumption that the volume change $\Delta V$ of a supercell with a defect can be determined in a similar manner as for a supercell containing bulk bcc-Fe and that the values of $\gamma$ are identical for both types of supercells. This procedure can be considered as a first approximation. In the following a more accurate (quasi-harmonic) approach is described briefly. However, since it requires a very high computational effort it is not employed here. For given temperatures the vibrational free energy may be calculated for different values of the supercell volume. Since the pressure is the negative derivative of the free energy with respect to the volume, the zero-pressure case corresponds to the minimum of the vibrational free energy with respect to the supercell volume. In this manner the frequencies and the corresponding vibrational free energy may be determined at given temperature, for the supercells with and without the defect. From these results defect free energies may be determined using Eq. (13).

In order to validate the computational method used in this work the calculated vibrational free energy $F^{vib}(T)$ and entropy $S^{vib}(T)$ of bulk bcc-Fe, fcc-Cu, fcc-Al, and hcp-Y were compared with data from literature, taking into account recent developments and discussions related to the SGTE database.[28,79,81-84] In general a very good agreement is found. For details the reader is referred to the Supplemental Material.[80]

### B. Influence of defects on vibrational frequencies

If a vacancy or a foreign atom is introduced into the bcc-Fe supercell under CV conditions the interatomic distances in the environment of the defect are increased or reduced and there is a buildup of a total internal pressure in the supercell (cf. Table III). The changed equilibrium atomic positions lead to a modification of the atomic force constants. This effect and the fact that the mass of the foreign atom is different to the Fe atomic mass cause the modification of the vibrational frequencies of the supercell. On the other hand, if the defect is introduced under ZP conditions not only the equilibrium positions of atoms in the defect environment are changed but there is also a position change of all the other atoms due to the increase or decrease of the supercell volume (cf. Table III). Both effects cause a change of force constants and lead, together with the mass effect, to an alteration of the vibrational frequencies of the supercell.

Figs. 1-3 depict the relative deviation of the vibrational frequencies $\omega_i$ of a supercell with a defect from the phonon frequencies $\omega_i^0$ of the bulk bcc-Fe supercell. The straight lines were obtained by a linear fit to the data points and shall only show the trends. The representation of $(\omega_i - \omega_i^0)/\omega_i^0$ as function of $\omega_i^0$ was found to be more suitable to illustrate frequency modifications than depicting the phonon density of states. Note that in all examples considered the numbering of $\omega_i^0$ and $\omega_i$ by the index $i$ starts at the highest frequency. In the case of the vacancy and the O interstitial where the total number of atoms in the supercell is not identical to that in the perfect bcc-Fe supercell the 3 lowest frequencies of the respective cell with the higher number of atoms were discarded. This does not affect the qualitative discussion on the modification of the vibrational frequencies due to the presence of a defect. Fig. 1 (a) shows that the presence of the vacancy leads to a relative decrease of most of the phonon frequencies. If the atomic positions were obtained by CV the relative decrease of the vibrational frequencies is slightly higher than in the case of ZP. The effect found for the vacancy in a supercell with 128 bcc sites is qualitatively similar to that for the smaller cell [Fig. 1 (b)]. However, the relative decrease of the vibrational frequencies is generally smaller since the effect of the vacancy on the whole phonon spectrum of the larger cell is weaker. Figs. 2 (a) and (b) depict the results for Cu in bcc-Fe for two different supercell sizes. The presence of Cu leads predominantly to a relative decrease of the vibrational frequencies. The relation between the data for the larger and smaller supercell is similar to the case of the vacancy. Present results for Cu are in accord with findings of Reith *et al.*[6] Fig. 3 (a) illustrates the case of Y in a supercell with 54 bcc sites. Here a relative decrease of the vibrational frequencies in the low frequency range and an increase in the high frequency range are observed if the equilibrium positions of atoms were obtained by CV. The vibrational frequencies mainly decrease if the atomic positions were obtained by ZP. For the O octahedral interstitial the relative increase of the vibrational frequencies dominates if the equilibrium

positions of atoms were determined by CV [Fig. 3 (b)]. The ZP-based results show frequency shifts in both directions. In the case of the O interstitial there exist three high frequencies with energies of about 0.096 (0.089), 0.047 (0.045), and 0.042 (0.042) eV, for CV- (ZP)-based calculations [not completely shown in Fig. 3 (b)]. Fig. 3 (c) depicts the relative deviation of the vibrational frequencies for the O-v pair. The behavior is qualitatively similar to that observed for the O interstitial. Three of the highest vibrational frequencies increase essentially, by about 0.025, 0.012, and 0.012 eV [not shown in Fig. 3 (c)].

After showing these illustrative examples two aspects will be discussed: (i) The change of the vibrational frequencies due to the introduction of the defect in the perfect crystal under CV conditions, and (ii) the difference between the phonon frequencies in the supercell with the defect in the CV case and those in the ZP case. In order to understand the change of the vibrational frequencies due to the presence of a defect the relaxation of the neighboring Fe atoms and the modification of the force constants is discussed in the CV case. For details the reader is referred to the Supplemental Material.[85] The presence of the vacancy causes a significant relaxation of Fe atoms in the first neighbor shell towards the vacancy position, whereas the opposite effect is observed for Cu. In both cases the relaxation in the second and every subsequent shell is decreasing but oscillates from shell to shell. Interestingly, the relaxation leads to a decrease of force constants of Fe atoms in the first neighbor shell of both the vacancy and the Cu substitutional atom. Furthermore, the force constants of the Cu atom are smaller than those of perfect bcc-Fe. On the average atoms in the second neighbor shell of v and Cu have force constants close to values for perfect Fe. From these findings one can conclude that the relative decrease of most of the vibrational frequencies observed for v and Cu in the CV case (cf. Figs. 1 and 2) is due to the decrease of force constants (since the frequency is proportional to the square root of the force constant) which is caused by the change of interatomic distances around the defect. The fact that the mass of Cu is slightly higher than that of Fe may also contribute to this effect. For the explanation of the difference between the vibrational frequencies of the supercell with the defect in the CV case and those in the ZP case, it is examined whether the relation between both sets of data can be approximately described according to an expression similar to Eq. (16)

$$\omega_i^{ZP} \approx \omega_i^{CV}\left(1 - \gamma' \frac{V - V_0}{V_0}\right) \qquad (17)$$

with data for $V_0$ and $(V - V_0)$ from Table III. Indeed, if the difference between $\omega_i^{ZP}$ and $\omega_i^{CV}$ is determined for each $i$, one obtains average values of $\gamma'$ between 1.65 and 1.88 (see Supplemental Material[86]) which are close to the phonon Grüneisen parameter of bcc-Fe.[87,88] An average over all data gives the value $\gamma' = 1.74$. Therefore, the difference between the CV- and ZP-based frequencies is predominantly due to the volume change of the whole supercell in the ZP case, while the relative arrangement of atoms in the environment of the defects is nearly identical in the two cases. This is an important result which is used in the discussion of the difference between CV- and ZP-based data for defect free energies (Sec. III C 4). Note that it is not self-evident that the relative arrangement of the Fe atoms around the defect is nearly the same for the CV and ZP results but this must be examined in each case considered. In the Supplemental Material[85] the relaxation of the Fe atoms in the first five neighbor shells of a defect is illustrated for the vacancy and the Cu atom. The trends obtained for the relative change of the force constants in the ZP and CV case (cf. Supplemental Material[85]) are consistent with the trend for the difference between $\omega_i^{ZP}$ and $\omega_i^{CV}$.

### C. Vibrational contributions to free formation and binding energies

#### *1. Vacancy and Cu*

First, the results for the single vacancy and the single Cu atom in bcc-Fe are presented since in these cases the only literature data are available for comparison. Fig. 4 (a) shows the vibrational contribution to the free formation energy of the vacancy. For supercells with 54, 128, and 250 bcc sites the data are depicted by thin, dashed and thick lines, respectively. The difference between the curves for the three cell sizes is very small for the ZP-based data ($F_f^{\text{vib,ZP}}$) whereas a size effect is found for the CV-based data ($F_f^{\text{vib,CV}}$). This indicates that only in the first case calculations using a supercell with 54 lattice sites yield completely accurate results. The data of Lucas *et al.*[5] are also given in the figure. They were calculated using equilibrium positions of atoms obtained by CV, a supercell with 128 bcc sites, and $3 \times 3 \times 3$ $k$ points. For supercells with 54 and 128 lattice sites the agreement with the CV-based data calculated in the present work is very good. On the other hand, Fig. 4 shows that the results based on equilibrium positions of atoms obtained by ZP ($F_f^{\text{vib,ZP}}$) differ from the data based on CV ($F_f^{\text{vib,CV}}$). The difference is considerably higher than that of the formation energies $E_f$ (Table II). The reason for this will be discussed later (Sec. III C 4). Fig. 4 demonstrates that with increasing temperature the phonon contribution to the free formation energy leads to a considerable reduction of the total free formation energy of the vacancy, i.e. the thermodynamic stability of the vacancy increases, which also contributes to an increase of its equilibrium concentration [cf. Eq. (6)]. Note that the nonzero value of $F_f^{\text{vib}}$ at $T = 0$ is due to the zero point vibrations. The application of the quasi-harmonic correction [Eq. (16)] to ZP-based phonon frequencies of the supercell with the vacancy and to that with perfect bcc-Fe leads to values of the free formation energy, that are almost identical to the original ZP-based data shown in Fig. 4. The slope of the curves corresponds to the vibrational contribution to the formation entropy [cf. Eq. (12)]. At relatively high temperatures $S_f^{\text{vib}}$ is a constant and about $4.8 k_B$ and $3.0 k_B$ in the CV and the ZP case, respectively. Above 500 K a constant value for $S_f^{\text{vib}}$ is obtained for all examples depicted in Figs. 4-7. This can be explained by considering Eq. (10) in the high-temperature limit $\hbar\omega_i \ll k_B T$, in combination with Eqs. (12) and (13). Then $S_f^{\text{vib}}$ becomes independent of temperature and contains only the logarithmically averaged phonon frequencies (cf. e.g. Ref. 78). Obviously, in the examples of Figs. 4-7 such a relation is already valid at temperatures slightly above the Debye temperature of bcc-Fe. This is not surprising because for other physical properties a similar validity range is known for the high-temperature limit (cf. e.g. Ref. 78).

The thin, dashed and thick lines in Fig. 5 (a) depict the vibrational contribution to the free formation or substitutional energy of Cu, for supercells with 54, 128, and 250 bcc sites, respectively. There is only a small size effect suggesting that the choice of a cell with 54 lattice sites leads already to satisfactory results. For comparison the data of Reith *et al.*[6] are also shown. These authors used a supercell with 64 bcc sites and equilibrium positions of atoms obtained by ZP. As in the case of the vacancy the thermodynamic stability of the Cu substitutional atom increases with temperature, which contributes to an increase of the Cu solubility in bcc-Fe [cf. Eq. (6)]. There is a qualitative agreement between the data of Ref. 6 and the ZP-based results ($F_f^{\text{vib,ZP}}$) of the present work. The remaining deviations should be mainly related to the use of a non-cubic supercell in Ref. 6. Similar to Fig. 4 there is a non-negligible difference between the data calculated using equilibrium positions of atoms obtained by CV and ZP. On the other hand, the values of the formation energy $E_f$ determined using CV and ZP are nearly identical (Table II). Unlike the case of the vacancy, there is an outwards relaxation of the atoms around the Cu atom which leads to an increase of the volume of the supercell (Table III) in the ZP case. Obviously, this effect results in a stronger temperature dependence of the vibrational contribution to the free formation energy than in the CV-based case (see also Sec. III C 4). It must be mentioned that such a difference was also discussed qualitatively by Reith et al.[6] but these authors showed only their ZP-based data explicitly. Quasi-harmonic corrections were applied to the ZP-based free formation energy and slight modifications at temperatures above 700 K were found [Fig. 5 (b)].

It must be noticed that in Figs. 4, 5, and in the following figures the temperature ranges up to 1200 K which is above the temperature of the ferromagnetic-to-paramagnetic transition (1043 K) and also slightly above the temperature of the $\alpha$-to-$\gamma$-phase transition (1183 K) of iron. From the viewpoint of magnetism the presented data are only valid below the Curie temperature. On the other hand the temperature dependence of the spontaneous magnetization in the ferromagnetic state is not considered here. If not stated otherwise it is always assumed that the magnetization of bulk iron corresponds to its value at $T = 0$.

### *2. Y, Ti, Cr, Mn, Ni, V, Mo, Si, Al, Co, O*

The vibrational contribution to the free formation energy of several solutes is shown in Figs. 6 (a)-(k) by the thick lines. The results were calculated for a Fe supercell with 54 bcc sites using the equilibrium positions of atoms determined by CV and ZP. In general a significant difference between the CV-based ($F_f^{\text{vib,CV}}$) and the ZP-based ($F_f^{\text{vib,ZP}}$) data is found. With the exception of Si the temperature dependence is qualitatively similar to that shown in Fig. 5 for Cu, i.e. the data for $F_f^{\text{vib,CV}}$ are higher than those for $F_f^{\text{vib,ZP}}$. This should be also due to fact that in these cases ZP yields an increase of the supercell volume (see also Sec. III C 4). For Si the behavior is similar to that of the vacancy (Fig. 4) because of the lower volume in the case of ZP. In contrast to Figs. 4 and 5 in many cases the CV- and ZP-based curves show opposite trends: While the latter decrease mostly with temperature, the former show often an increase. In this case, the thermodynamic stability of the foreign atom in bcc-Fe decreases with temperature. Such a behavior was also found in the few recent papers on defect free energies, i.e. in the work of Reith et al.[6] for the pair of Fe atoms in fcc-Cu, in the work of Bjørheim et al.[14] for the oxygen vacancy in ZnO, in the paper of Gryaznov et al.[15] for the oxygen vacancy in Sr-doped complex pervoskites, and in the work of Kosyak et al.[16] for some point defects in $Cu_2ZnSnS_4$. The difference between $F_f^{vib,CV}$ and $F_f^{vib,ZP}$ is highest for Y and O, e.g. about 0.5-0.6 eV at 1000 K. In both cases the corresponding formation energies $E_f^{CV}$ and $E_f^{ZP}$ differ only by about 0.2 eV (Table II). On the other hand, the difference between the values of the formation energy of Si determined by CV and ZP is only about 0.0003 eV which might be the reason for the small difference of the corresponding curves for $F_f^{vib,CV}$ and $F_f^{vib,ZP}$ in Fig. 6 (h) (about 0.019 eV at 1000 K). In the other cases the difference $E_f^{CV} - E_f^{ZP}$ is between 0.04 and 0.003 eV, whereas $F_f^{vib,CV}$ and $F_f^{vib,ZP}$ show larger differences, e.g. between 0.1 and 0.2 eV at 1000 K.

The application of quasi-harmonic corrections to the ZP-based free formation energy was investigated for Y, Ti, Ni, V, Mo, Al, and Co. For Y, Ni, and Co the modifications obtained are negligible. In the other cases the corresponding curves are shown in the figures. In general the change of the corresponding values of the free formation energy is small and only visible above about 700 K. A stronger modification is found for Al. The different results can be explained by the interplay of the three terms in Eq. (13) which depends on the particular foreign atom considered. The shift of the vibrational frequencies according to Eq. (16) was determined using values of the Grüneisen parameter which were obtained from literature data for the volume expansion coefficient, the bulk modulus, the density, and the specific heat capacity.[87-90] Details are given in the Supplemental Material.[80,91,92]

### 3. O-v pair

The temperature dependence of the vibrational contribution to the free formation energy of the O-v pair depicted in Fig 7 (a) looks somewhat different to that of single foreign atoms. The difference between the curves for $F_f^{\text{vib,CV}}$ and $F_f^{\text{vib,ZP}}$ is about 0.1 eV at 1000 K which is substantially smaller than in the case of the O interstitial. This trend is similar to that found for the difference between the corresponding formation energies (Sec. II C). Fig 7 (b) illustrates the vibrational contribution to the free binding energy of the O-v pair. The CV- and ZP-based data differ considerably, e.g. 0.3 eV at 1000 K. Such differences should be generally expected for the free binding energy of nanoclusters consisting of point defects and foreign atoms. At 1000 K $F_{bind}^{\text{vib,ZP}}$ is about 0.34 eV while the ground state value $E_{bind}^{ZP}$ is -1.53 eV (Sec. II D). Neglecting $F_{bind}^{\text{vib,ZP}}$ would therefore lead to an error of the total free binding energy of about 20%.

### 4. Relation between $F_f^{\text{vib,CV}}$ and $F_f^{\text{vib,ZP}}$

Summarizing the results depicted in Figs. 4-7 one can conclude that the vibrational contribution to the free formation and binding energy is generally not negligible. In the examples considered, the ratio $\left|F_f^{\text{vib}}(T=1000\,\text{K})\right|/\left|E_f\right|$ ranges between 0.01 and 8. The highest number is obtained for Mo because this solute has a very low formation energy $E_f$. In most cases $F_f^{\text{vib,ZP}}$ decreases monotonously with temperature, which may lead to an increase of the equilibrium concentration or solubility of the defects [cf. Eq. (6)], whereas for $F_f^{\text{vib,CV}}$ both an increase and a decrease is found. The consideration of the quasi-harmonic correction [Eq. (16)] leads only to a minor modification of this picture. On the other hand, the corrected ZP-based data are valid under zero-pressure conditions at higher temperatures than the data obtained within the purely harmonic approach. These conditions are usually realized in experiments. Vibrational effects must be also considered in the determination of the total free binding energy of defect clusters as illustrated in the case of the O-v pair. Another important finding concerns the origin of the difference between CV- and ZP-based data. It is

D. Murali, M. Posselt, M. Schiwarth: First-principles calculation of defect free energies… MS#BD12759        **25**

is mainly due to the volume change of the supercell while the relative arrangement of atoms in the environment of the defects is nearly identical in the two cases.

Using Eq. (17) with the vibrational frequencies $\omega_i^{CV}$, as well as $V_0$ and $(V-V_0)$ from Table III, and $\gamma' = 1.74$ (cf. Sec. III B) yields data for $F_f^{vib,ZP}(T)$ [via Eqs. (10) and (13)] which are almost identical to those directly determined by the frequencies $\omega_i^{ZP}$. This holds for all cases considered in Figs. 4-7. The result is not surprising since the vibrational frequencies are the most important quantities entering the expressions for the vibrational free energy, and it was shown in Sec. III B that $\omega_i^{CV}$ can be really used to calculate $\omega_i^{ZP}$ approximately by Eq. (17). One can therefore conclude that, if a frequency transformation according to Eq. (17) is possible, separate calculations of $\omega_i^{CV}$ and $F_f^{vib,CV}$ on the one hand, and of $\omega_i^{ZP}$ and $F_f^{vib,ZP}$ on the other hand, are not necessary. The above discussion is somewhat similar to that related to the transformation (9) between the formation energies $E_f^{CV}$ and $E_f^{ZP}$ in the ground state. In the latter case results of CV and ZP fully converge for sufficiently large supercells (cf. Sec. II B). How is the situation in the case of $F_f^{vib,CV}$ and $F_f^{vib,ZP}$? To clarify this question an analytical expression is derived for the difference between $F_f^{vib,CV}$ and $F_f^{vib,ZP}$. Since the second term in Eq. (17) is usually relatively small, i.e. $(V-V_0)/V_0 \ll 1$, one obtains

$$F_f^{vib,ZP}(T) - F_f^{vib,CV}(T) = -\gamma' \frac{V-V_0}{V_0}\left[\sum_i \frac{1}{2}\hbar\omega_i^{CV} + \sum_i \hbar\omega_i^{CV}\left(e^{\frac{\hbar\omega_i^{CV}}{k_B T}} - 1\right)^{-1}\right] \quad (18)$$

Firstly, this expression clearly shows that $F_f^{vib,ZP} - F_f^{vib,CV}$ is positive if $V-V_0$ (cf. Table III) is negative and *vice versa*. This explains why $F_f^{vib,ZP} > F_f^{vib,CV}$ holds in the case of the vacancy and Si and why $F_f^{vib,ZP} < F_f^{vib,CV}$ is valid in all the other examples considered (Figs. 4-7). Secondly, Eq. (18) clearly shows that the difference $F_f^{vib,ZP} - F_f^{vib,CV}$ does not vanish for $N \rightarrow \infty$ since $(V-V_0)/V_0 \propto 1/N$ and the sum over $i$ is proportional to $N$. This is confirmed by Figs. 4 and 5 which do not show a decrease of the difference $F_f^{vib,ZP} - F_f^{vib,CV}$ with increasing size of the supercell. The comparison of $F_f^{vib,CV}(0) - F_f^{vib,ZP}(0)$ obtained from Figs. 4-7 with the corresponding values determined by Eq. (18) shows a good agreement. For details the reader is referred to the Supplemental Material.[92] According to Eq. (18) the absolute value of the difference $F_f^{vib,ZP}(T) - F_f^{vib,CV}(T)$ increases monotonously with temperature which is also observed in Figs. 4-7. The difference between the vibrational contributions to the formation entropy $S_f^{vib,ZP}$ and $S_f^{vib,CV}$ is obtained from Eqs. (18) and a relation similar to Eq. (12) (cf. Refs. 20 and 78)

$$S_f^{vib,ZP} - S_f^{vib,CV} = \gamma' C_V \frac{V-V_0}{V_0} \tag{19}$$

where $C_V = (\partial U^{vib}/\partial T)_V$ is the heat capacity at constant volume (phonon contribution, cf. Ref. 78). Since $\gamma'$ was found to be nearly equal to the phonon Grüneisen parameter of bcc-Fe, this quantity can be also expressed by other material parameters using the know relation

$$\gamma' = \frac{\beta B_0}{C_V} V_0 \tag{20}$$

where $\beta$ is the volume expansion coefficient (phonon contribution, cf. Ref. 78). Then Eq. (19) simplifies to

$$S_f^{vib,ZP} - S_f^{vib,CV} = \beta B_0 (V-V_0) \tag{21}$$

In the Supplemental Material[92] the difference $S_f^{vib,ZP} - S_f^{vib,CV}$ obtained from the constant slopes of $F_f^{vib,ZP}(T)$ and $F_f^{vib,CV}(T)$ in the temperature range above 500 K (Figs. 4-7) is compared with that determined by Eq. (21) and a good agreement is found. This demonstrates once again that CV- and ZP-based results are fully consistent with each other. Note that an expression similar to Eq. (21) is also valid for the difference between the total formation entropy at zero pressure and constant volume in the case of intrinsic defects (cf. Refs. 20 and 93). It can be obtained by general thermodynamic considerations, also under the assumption $(V-V_0)/V_0 \ll 1$.

**D. Vibrational contributions to the free migration energy and the diffusivity of vacancies**

The thick solid lines in Fig. 8 depict the vibrational contribution to the free migration energy $\Delta F_{mig}^{vib}(T)$ of the vacancy as function of temperature. $\Delta F_{mig}^{vib}(T)$ shows a non-monotonous behavior. As in the case of the vibrational contribution to the free formation energy $F_f^{vib}(T)$, there is a difference between the results obtained using atomic positions determined by CV and those by ZP although the corresponding values of $E_{mig}$ are equal. However, while the CV-based values of $F_f^{vib}(T)$ are higher than the ZP-based data (Fig. 4) the opposite order is found for $\Delta F_{mig}^{vib}(T)$. The application of the quasi-harmonic correction [Eq. (16)] to the ZP-based phonon frequencies causes only a slight modification of the values for $\Delta F_{mig}^{vib}(T)$ at temperatures above 700 K.

## IV. EFFECT OF ELECTRONIC EXCITATIONS

### A. Calculation details

The contribution of electronic excitations to the free energy is defined by (cf. Refs. 13, 23, 24, 78, and 94).

$$F^{el}(T) = U^{el}(T) - T S^{el}(T) \tag{22}$$

The corresponding internal energy is given by

$$U^{el}(T) = \sum_{i=1}^{2} \left[ \int_{\varepsilon_F}^{\infty} n_i(\varepsilon)(\varepsilon - \varepsilon_F) f(\varepsilon, T, \mu) d\varepsilon + \int_{0}^{\varepsilon_F} n_i(\varepsilon)(\varepsilon_F - \varepsilon)[1 - f(\varepsilon, T, \mu)] d\varepsilon \right] \tag{23}$$

where $n(\varepsilon)$ and $f(\varepsilon,T,\mu)$ are the electronic density of states (DOS) and the Fermi-Dirac distribution function, respectively, while $\varepsilon_F$ stands for the Fermi energy. In Eq. (23) the sum is over the two spin orientations. The quantity $\mu(T)$ denotes the chemical potential of the electrons which is implicitly given by

$$N^{el} = \sum_{i=1}^{2} \int n_i(\varepsilon) f(\varepsilon,T,\mu) d\varepsilon \qquad (24)$$

with $N^{el}$ as the total number of electrons considered. The electronic entropy $S^{el}(T)$ is defined by

$$S^{el}(T) = -k_B \sum_{i=1}^{2} \int \left[ f(\varepsilon,T,\mu) \ln f(\varepsilon,T,\mu) + (1-f(\varepsilon,T,\mu)) \ln(1-f(\varepsilon,T,\mu)) \right] n_i(\varepsilon) d\varepsilon. \qquad (25)$$

In the temperature range below about the half Debye temperature (200 - 250 K) the effect of electron-phonon interaction may influence the value of the free energy. This was investigated using the expressions given by Grimvall[78] with the corresponding data for the electron-phonon interaction parameter. It was found that the consideration of this effect does not lead to any visible changes in the representation of the free energy shown in this work.

The influence of electronic excitations on the free energy of bulk bcc-Fe and the other bulk reference materials considered in this work is smaller than that of phonon excitations. For characteristic examples this is illustrated in the Supplemental Material.[80] The electronic contribution to the free formation energy $F_f^{el}(T)$ of the vacancy, of foreign atoms, and of the O-v pair, as well as the electronic contribution to the free binding energy of the O-v pair $F_{bind}^{el}(T)$ are defined by relations similar to Eqs. (2-5).

### B. Electronic contribution to free formation and binding energies

The sum of vibrational and electronic contributions to the free formation energy, $F_f^{vib}(T) + F_f^{el}(T)$, is depicted by the thin solid lines in Figs. 4-7, for a supercell with 54 bcc sites. In contrast to $F_f^{vib}(T)$ there is almost no difference between the CV- and ZP- based values of $F_f^{el}(T)$. The only exceptions are Ni and Y where slight differences occur. Although the electronic contributions to the free formation energy are smaller than the vibrational ones they are generally not negligible. The ratio $|F_f^{el}(T=1000\,\text{K})|/|E_f|$ varies between 0.01 and 4.0 where the highest number is again obtained for Mo. Depending on the defect considered the electronic contributions can amplify the deviation of the total free energy from the ground state value, which are caused by phonon excitations, or can compensate it. For v, Y, O, and the O-v pair $F_f^{el}(T)$ is very small compared to $F_f^{vib}(T)$ whereas it cannot be neglected in the other cases. The largest values (>0.04eV at 1000 K) were found for Al, Mo, and V substitutional atoms. The electronic contribution to the free binding energy of the O-v pair is insignificant. The values of $F_f^{el}(T)$ show an increase (Co, Cr, Mn, Mo, Ni, O, O-v, Ti, V, v, Y) or a decrease (Al, Cu, Si) versus temperature. Such qualitative differences were also found by Satta et al.[23,24] who considered the corresponding formation entropy $S_f^{el}(T)$: While in the case of the vacancy in bcc-W $S_f^{el}(T)$ increases with $T$ a decrease was found for the vacancy in bcc-Ta. The explanation for such a behavior can be given considering the Sommerfeld approximation. For low temperatures ($k_B T \ll \mu$) Eqs. (22-25) reduce to (cf. Refs. 78 and 94)

$$F^{el}(T) = -\frac{\pi^2}{6}(k_B T)^2 \sum_{i=1}^{2} n_i(\varepsilon_F) \qquad (26)$$

Therefore, the increase or decrease of $F_f^{el}(T)$ depends essentially on the sign of the difference between the values of the density of states at the Fermi energy in the system with the defect and in the corresponding reference systems, since $F_f^{el}(T)$ is given by an expression similar to Eqs. (2-4).

**C. Electronic contribution to the free migration energy and the diffusivity of vacancies**

The electronic contribution to the free migration energy of the vacancy is calculated by an expression similar to Eq. (14). The CV- and ZP-based results are nearly identical and very small compared to $\Delta F_{mig}^{vib}(T)$. The thin lines in Fig. 8 show the sum $\Delta F_{mig}^{vib}(T) + \Delta F_{mig}^{el}(T)$. The replacement of $\Delta F_{mig}^{vib}(T)$ by $\Delta F_{mig}^{vib}(T) + \Delta F_{mig}^{el}(T)$ in Eq. (15) does not lead to any significant changes of the vacancy diffusivity. Therefore one can conclude that the influence of electronic excitations on both the total free migration energy and the total free formation energy (cf. Sec. IV B) of the vacancy is negligible.

## V. SELF-DIFFUSION IN BCC-FE

Since in bcc-Fe self-diffusion occurs predominantly via the vacancy mechanism the product of the diffusivity $D(T)$ of vacancies and their equilibrium concentration $C_{eq}(T)$ [cf. Eq. (6)] leads to the self-diffusion coefficient

$$D_{SD}(T) = f\, D(T)\, C_{eq}(T) \tag{27}$$

where $f = 0.727$ is the correlation factor.[95] The temperature dependence of $C_{eq}(T)$ and $D(T)$ is illustrated in the Arrhenius plot shown in Fig. 9. Results of previous sections (Secs. IV B and C) show that contributions of electronic excitations to $C_{eq}(T)$ and $D(T)$, and therefore to self-diffusion, can be neglected. In Fig. 9 lines obtained by CV and ZP have different absolute values but are practically parallel. Since the ratio of the CV- and ZP-based data of $C_{eq}(T)$ is nearly equal to the reciprocal of the corresponding ratio of the data for $D(T)$, the CV- and ZP-based self-diffusion coefficients $D_{SD}(T)$ are almost the same in the Arrhenius plot depicted in Fig. 10. The self-diffusion coefficient can be also expressed by

$$D_{SD}(T) = D_0 \exp\left[-\frac{E_{SD}^{eff}}{k_B T}\right] \tag{28}$$

The inset of Fig. 10 shows the values of the effective activation energy $E_{SD}^{eff}$ and the prefactor $D_0$. The DFT data of Huang *et al.*[7] obtained by calculations that were very similar to those performed in the present work under CV conditions lie completely within the range of the other lines depicted in Fig. 10. The value of $E_{SD}^{eff}$ obtained from the total free formation and migration energy of the vacancy is only slightly different to the corresponding value in the ground state given by $E_{SD} = E_f + E_{mig} = 2.84$ eV. This is not self-evident but due to the fact, that both in the CV and the ZP case the vibrational contributions affect the free formation and migration energy of the vacancy in an opposite manner (cf. Fig. 9).

All DFT results hitherto discussed are valid for bcc-Fe with maximum magnetization. For a detailed comparison with experimental self-diffusion data the temperature dependence of the magnetization $M(T)$ must be taken into account. For this purpose the well-established semi-empirical model originally proposed by Ruch *et al.*[96], cf. also Refs. 7, 97, and 98, is employed in the present work. The model relates the self-diffusion coefficient in the ferromagnetic state with a magnetization $M(T)$ to the parameters $E_{SD}^{pm}$ and $D_0^{pm}$ determined in the paramagnetic state

$$D_{SD}^{fm}(T) = D_0^{pm} \exp\left[-\frac{E_{SD}^{pm}(1+\alpha s^2)}{k_B T}\right] \tag{29}$$

$$D_0^{fm} = D_0^{pm} = D_0 \tag{30}$$

where $s(T) = M(T)/M(0)$. In the case of the completely ordered ferromagnetic state ($s=1$) the corresponding parameters are given by

$$E_{SD}^{fm}(s=1) = E_{SD}^{pm}(1+\alpha) = E_{SD}^{eff}. \tag{31}$$

This leads to

$$E_{SD}^{fm}(s) = E_{SD}^{eff} \frac{(1+\alpha s^2)}{(1+\alpha)} \tag{32}$$

For bcc-Fe the parameter $\alpha$ is equal to 0.156 (Refs. 7, 97, and 98). In the present work the data from the experimental study of Crangle et al.[99] were used for $s(T)$. Using the DFT data for $E_{SD}^{eff}$ and $D_0$ as well as the above considerations related to $M(T)$ the self-diffusion coefficient both in the paramagnetic and in the ferromagnetic range can be determined by

$$D_{SD}^{mag}(T) = D_0 \exp\left[-\frac{E_{SD}^{fm}(s)}{k_B T}\right]. \tag{33}$$

Due to the magnetic phase transition a non-Arrhenius behavior is observed. Fig. 11 demonstrates that the calculated curves show a very good agreement with the most reliable experimental self-diffusion data from literature.[100-102] Furthermore, a good agreement with the theoretical results of Huang et al.[7] is found (not shown) who employed only CV-based data and a calculation procedure very similar to that performed in the present work. It must be emphasized that the good agreement between the CV- and ZP-based curves found in the present work is accidental and cannot be generalized to other cases of self-diffusion. That also means that the very good agreement of the theoretical results with experimental data found in Ref. 7 is not self-evident. Note that in the case of the vacancy quasi-harmonic corrections to the ZP-based results were found to be very small (Secs. III C 1 and III D), so that, finally, both the CV- and ZP-related self-diffusion data presented above are valid in the whole temperature range and for zero-pressure conditions, which are usually realized in experiments.

A full first-principle calculation of self-diffusion in bcc-Fe must include contributions of magnon excitations to the free formation and migration energy of the vacancy. Although a comprehensive treatment of this problem does not exist yet, attempts to solve this issue have been made. Ding et al.[103] have recently obtained the value of the parameter $\alpha$ by DFT calculations on paramagnetic bcc-Fe with and without a vacancy. Very recently Sandberg et al.[104] attempted to calculate the magnetic contribution to the activation energy of self-diffusion using a combination of DFT and Heisenberg model Monte Carlo simulations.

# VI. SUMMARY AND CONCLUSIONS

The ground state energetics of the vacancy, of numerous foreign atoms and of the oxygen-vacancy pair in bcc-Fe was determined using supercells with 54, 128, and 250 lattice sites. In every case considered two types of calculations were performed: After introduction of a defect into the supercell: (i) the positions of atoms are relaxed but volume and shape of the supercell are held constant (constant volume calculations – CV) and (ii), additionally, volume and shape of the supercell are relaxed so that the total pressure on the supercell is zero (zero pressure calculations - ZP). In many cases the defect energetics determined by the two methods is nearly equal and in good agreement with DFT data from literature. The remaining differences between results of CV and ZP can be explained within the framework of an elasticity approach using the data for the internal pressure in the supercell obtained by CV and for the volume change determined by ZP. For sufficiently large supercells these differences must tend towards zero.

The vibrational contribution to the free energy of the defects was calculated within the framework of the harmonic approximation using the equilibrium atomic positions determined by both CV and ZP. In most cases the ZP-based free formation energy decreases monotonously with temperature, which may contribute to an increase of the equilibrium concentration or solubility of the defects, whereas for CV-based data both an increase and a decrease were found. The application of a quasi-harmonic correction to ZP-based data does not modify this picture significantly. However, the corrected data are valid under zero-pressure conditions at higher temperatures than in the framework of the purely harmonic approach. These conditions are usually realized in experiments. In the examples considered, at 1000 K the absolute value of the ratio of the vibrational contribution to the free formation energy to the formation energy in the ground state ranges between 0.01 and 8.0. Vibrational effects must be also considered in the determination of the total free binding energy of defect clusters as illustrated in the case of the O-v pair. Another important finding concerns the origin of the difference between CV- and ZP-based free energy data. It is mainly due to the volume change of the supercell while the relative arrangement of atoms in the environment of the defects is nearly identical in the two cases. This allows a simple transformation similar to the quasi-harmonic approach. As a consequence of this finding a separate calculation of CV- and ZP-based phonon frequencies and of the corresponding defect free energies is not necessary. Based on these results it was demonstrated analytically that the values of CV- and ZP-based free energies do not become equal with increasing supercell size. This is in contrast to the size convergence found for ground state energetics

For the defects studied in this work, at 1000 K the absolute value of the ratio of the electronic contribution to the free formation energy to the formation energy in the ground state ranges between 0.01 and 4.0. Unlike the phonon case, the electronic contribution does not significantly depend on whether CV- and ZP-based equilibrium positions of atoms are employed in the calculations. Depending on the defect considered the electronic contribution can amplify the deviation of the total free energy from the ground state value due to the contribution of phonon excitations, or can compensate it. The results of the present work yield a variety of temperature dependencies for the total defect free energy. In some cases this behavior is similar to that obtained recently by considering the influence of anharmonic vibrational effects on the free formation energy of the vacancy in Al and Cu.[11]

While the use of CV- and ZP-based equilibrium positions of atoms yields the same value for the migration barrier of the vacancy in the ground state, there is some difference between the corresponding data for the free migration energy. The ratio of the respective CV- and ZP-based results for the vacancy diffusivity is nearly equal to the reciprocal of that for the data of the equilibrium concentration. This yields almost identical CV- and ZP-based values for the self-diffusion coefficient. Obviously, this agreement is accidental and cannot be generalized to other cases of self-diffusion. The influence of electronic excitations on self-diffusion is negligible. In all DFT calculations it was assumed that the spontaneous magnetization corresponds to that in the ground state. In order to compare with experimental self-diffusion data a well-known semi-empirical model was employed that considers self-diffusion influenced by the temperature dependence of magnetization. The transformation led to self-diffusion data that are in very good agreement with those obtained in experiments. The temperature dependence of the magnetization is related to magnon excitations. A full first-principle treatment on the influence of this effect on the defect free formation and binding energies does not exist yet. The development of such a new method should be an important task for future research.

The results of this work are very important for multi-scale modeling of nanostructure evolution in ferritic Fe alloys at elevated temperatures using kinetic Monte Carlo simulations or rate theory. The data for the total free formation, binding and migration energy of defects determined in this work can be used to derive temperature-dependent parameters for the description of atomic interactions and diffusion jumps. These data can replace the presently employed input parameter values obtained from ground state calculations. In this manner the accuracy of modeling may be improved considerably. Since experiments are usually performed under zero pressure conditions the usage of the ZP-based free energy data corrected by the application of the quasi-harmonic approach is recommended for the parameterization. In order to extend the set of temperature dependent input parameters the free energy of several relevant nanoclusters should be also determined by DFT using the method outlined in this work. The general approach employed in the present study can be also utilized to determine defect free energies in other solids.

**ACKNOWLEDGEMENTS**

This work also contributes to the Joint Programme on Nuclear Materials (JPNM) of the European Energy Research Alliance (EERA).

TABLE I. Calculated lattice parameters in comparison with experimental data. The supercell sizes, numbers of $k$ points, and the pseudopotentials used in the present work are given as well. The superscripts correspond to the numbers given in the list of references.

| Solid | Lattice parameters | | | | Supercells and $k$ points used in this work | | PAW-PBE pseudopotentials (VASP notation) |
| --- | --- | --- | --- | --- | --- | --- | --- |
| | This work | | Experimental data | | | | |
| | $a_0$ (Å) | $c_0$ (Å) | $a_0$ (Å) | $c_0$ (Å) | Lattice sites | $k$ points | |
| Fe (bcc) | 2.83359 | | 2.86[50] | | 54 | 6×6×6 | Fe |
| | 2.83396 | | | | 128 | 3×3×3 | |
| | 2.83311 | | | | 250 | 2×2×2 | |
| Cu (fcc) | 3.63465 | | 3.61[51] | | 32 | 6×6×6 | Cu |
| Y (hcp) | 3.64740 | 5.73060 | 3.64[52] | 5.73[52] | 54 | 3×3×3 | Y_sv |
| Ti (hcp) | 2.92075 | 4.63279 | 2.95[53] | 4.68[53] | 54 | 3×3×3 | Ti_pv |
| Cr (bcc) | 2.86470 | | 2.91[54] | | 54 | 6×6×6 | Cr_pv |
| Mn (cubic) | 8.55680 | | 8.91[55] | | 58 | 6×6×6 | Mn_pv |
| Ni (fcc) | 3.51766 | | 3.52[56] | | 32 | 8×8×8 | Ni |

| | | | | | | |
|---|---|---|---|---|---|---|
| **V (bcc)** | 2.99681 | | 3.03[57] | | 54 | 6×6×6 | V_sv |
| **Mo (bcc)** | 3.16305 | | 3.15[54] | | 54 | 6×6×6 | Mo_sv |
| **Si (diamond)** | 5.46776 | | 5.43[58] | | 54 | 6×6×6 | Si |
| **Al (fcc)** | 4.04089 | | 4.05[59] | | 32 | 8×8×8 | Al |
| **Co (hcp)** | 2.49167 | 4.02501 | 2.50[60] | 4.06[60] | 54 | 6×6×6 | Co |
| **FeO (NaCl)** | 4.30921 | | 4.32[61] | | 8 | 12×12×12 | Fe, O |

TABLE II. Formation energy data determined under constant volume (CV) and zero pressure (ZP) conditions for three different supercell sizes, in comparison with DFT data from literature. The superscripts correspond to the numbers given in the list of references.

| Defect | $E_f$ (eV) | | | | | | | |
|---|---|---|---|---|---|---|---|---|
| | This work (CV) | | | *Literature data* | This work (ZP) | | | *Literature data* |
| | 54 | 128 | 250 | | 54 | 128 | 250 | |
| v | 2.1652 | 2.1605 | 2.2036 | 2.16[5], 2.15…2.17[34], 2.23[7], 2.23[46], 2.18…2.20[2] | 2.1598 | 2.1584 | 2.2005 | 1.95-2.01[43], 2.18… 2.21[2] 2.18[48], 2.15…2.17[34] |
| Cu | 0.73406 | 0.73190 | 0.73675 | 0.78[6], 0.76[32], 0.54…0.55[43], 0.55[44], 0.52[2] | 0.72850 | 0.72852 | 0.73633 | 0.77[6], 0.50…0.55[43] |
| Y | 2.2000 | 2.1289 | | 2.02[40], 2.01[41], 1.86[49] | 2.0012 | 2.0333 | | |
| Ti | -0.78166 | -0.78804 | | -0.80[40], -0.92[41], -0.79[32], -0.8[49] | -0.79742 | -0.79741 | | |
| Cr | -0.27194 | -0.25816 | | -0.11[32], -0.3[49] | -0.27938 | -0.26211 | | |
| Mn | 0.19436 | 0.20987 | | 0.25[32], -0.10…-0.14[44], -0.16[45] | 0.18676 | 0.20507 | | |
| Ni | 0.096970 | 0.09130 | | 0.13[32], -0.12…-0.22[44], -0.17[45] | 0.085616 | 0.08529 | | |
| V | -0.73849 | -0.73357 | | -0.70[32] | -0.74566 | -0.73822 | | |

| | | | | | | | |
|---|---|---|---|---|---|---|---|
| **Mo** | 0.024591 | 0.031063 | | | -0.012549 | 0.012953 | |
| **Si** | -1.1986 | -1.1649 | -1.08…-1.12[44] | | -1.1989 | -1.1649 | |
| **Al** | -0.74196 | -0.75064 | | | -0.74541 | -0.75381 | -0.93[47] |
| **Co** | -0.13627 | -0.13464 | -0.11[32] | | -0.13784 | -0.13609 | |
| **O** | 1.4557 | 1.4328 | 1.41[40], 1.45[42], 1.35[49] | | 1.2435 | 1.33512 | |
| **O-v** | 1.8857 | 1.9271 | | | 1.8762 | 1.92058 | |

TABLE III. Internal pressure $p$ in the supercell obtained under constant volume (CV) conditions (lattice constant $a_0$ and supercell volume $V_0$) as well as the lattice parameter $a$ and the volume change $V - V_0$ determined under zero pressure (ZP) conditions. If CV yields a negative/positive pressure a volume decrease/increase is obtained in the case of ZP. Results for three different supercell sizes are shown. For the O interstitial and the O-v pair a tetragonal distortion is obtained by ZP. This effect is characterized by the parameters $a$ and $b$.

| Defect | CV | | | ZP | | | | | |
|---|---|---|---|---|---|---|---|---|---|
| | $V_0 = 614.29$ Å³ | $V_0 = 1456.7$ Å³ | $V_0 = 2842.5$ Å³ | $p = 0$ | | | | | |
| | $p$ (eV/Å³) | | | $a$ (Å) | | | $V - V_0$ (Å³) | | |
| | 54 | 128 | 250 | 54 | 128 | 250 | 54 | 128 | 250 |
| v | $-4.7751 \times 10^{-3}$ | $-2.2096 \times 10^{-3}$ | $-1.4783 \times 10^{-3}$ | 2.8295 | 2.8321 | 2.8319 | -2.6703 | -2.9247 | -3.5622 |
| Cu | $4.5129 \times 10^{-3}$ | $1.7852 \times 10^{-3}$ | $9.3399 \times 10^{-4}$ | 2.8373 | 2.8354 | 2.8339 | 2.4137 | 2.2656 | 2.2874 |
| Y | $2.7483 \times 10^{-2}$ | $1.1844 \times 10^{-2}$ | | 2.8557 | 2.8431 | | 14.506 | 14.667 | |
| Ti | $7.5028 \times 10^{-3}$ | $3.2081 \times 10^{-3}$ | | 2.8396 | 2.8366 | | 3.9393 | 3.9999 | |
| Cr | $5.1059 \times 10^{-3}$ | $1.9224 \times 10^{-3}$ | | 2.8377 | 2.8353 | | 2.6811 | 2.1183 | |

| | | | | | | | |
|---|---|---|---|---|---|---|---|
| **Mn** | $5.0497 \times 10^{-3}$ | $2.1221 \times 10^{-3}$ | 2.8378 | 2.8355 | 2.7566 | 2.3680 | |
| **Ni** | $5.9923 \times 10^{-3}$ | $2.3530 \times 10^{-3}$ | 2.8385 | 2.8359 | 3.2156 | 2.9247 | |
| **V** | $5.2869 \times 10^{-3}$ | $2.0597 \times 10^{-3}$ | 2.8377 | 2.8356 | 2.6540 | 2.5918 | |
| **Mo** | $1.1923 \times 10^{-2}$ | $4.8753 \times 10^{-3}$ | 2.8429 | 2.8375 | 6.0831 | 6.0503 | |
| **Si** | $-8.3642 \times 10^{-4}$ | $-2.30935 \times 10^{-4}$ | 2.8330 | 2.8338 | 0.4023 | 0.3072 | |
| **Al** | $3.8388 \times 10^{-3}$ | $1.6914 \times 10^{-3}$ | 2.8370 | 2.8355 | 2.2031 | 2.3231 | |
| **Co** | $2.3742 \times 10^{-3}$ | $1.04857 \times 10^{-3}$ | 2.8355 | 2.8347 | 1.2285 | 1.1968 | |
| **O** | $2.1877 \times 10^{-2}\,(a)$ $3.4132 \times 10^{-2}\,(b)$ | $9.1579 \times 10^{-3}\,(a)$ $1.4183 \times 10^{-2}\,(b)$ | $2.8403\,(a)$ $2.8826\,(b)$ | $2.8374\,(a)$ $2.8536\,(b)$ | 13.582 | 13.688 | |
| **O-v** | $4.3080 \times 10^{-3}\,(a)$ $6.6276 \times 10^{-3}\,(b)$ | $1.6565 \times 10^{-3}\,(a)$ $2.7523 \times 10^{-3}\,(b)$ | $2.8350\,(a)$ $2.8440\,(b)$ | $2.8340\,(a)$ $2.8382\,(b)$ | 2.8496 | 2.7993 | |

# Figure captions

Fig. 1 (Color online) Relative deviation of the vibrational frequencies $\omega_i$ of the supercell with the vacancy from the vibrational frequencies $\omega_i^0$ of the perfect bcc-Fe supercell. The blue and red symbols show CV- and ZP-based results, respectively. The straight lines were obtained by a linear fit to the data points and shall illustrate the trends. Data calculated for supercells with 54 (a) as well as 128 bcc sites (b) are depicted.

Fig. 2 (Color online) The quantity $(\omega_i - \omega_i^0)/\omega_i^0$ for the Cu solute in bcc-Fe. Results for supercells with 54 (a) and 128 (b) bcc sites are presented. The general style of the presentation is explained in the caption of Fig. 1.

Fig. 3 (Color online) The quantity $(\omega_i - \omega_i^0)/\omega_i^0$ for Y (a), O (b), and the O-v pair (c) in bcc-Fe. A supercell containing 54 bcc sites was used in the calculations. The style of the presentations is identical to that in Figs. 1 and 2.

Fig. 4 (Color online) Vibrational contribution to the free formation energy of a single vacancy in bcc-Fe versus temperature, determined for supercells with 54 (thin lines), 128 (dashed lines) and 250 (thick lines) lattice sites (a). Note that the thin, dashed and thick red lines are almost identical. For a supercell with 54 bcc sites the vibrational contribution (thick lines) as well as the sum of vibrational and electronic contributions to the free energy (thin lines) is shown in (b). The thin and thick lines are not very different. In both figures the CV-based and the ZP-based free formation energies are depicted by blue and red color, respectively. The DFT results of Lucas *et al.*[5] are represented by black symbols.

Fig. 5 (Color online) Vibrational contribution to the free formation energy of a single Cu atom in bcc-Fe calculated for supercells with 54 (thin lines), 128 (dashed lines), and 250 (thick lines) lattice sites (a). In this representation the blue dashed line and the blue thick line cannot be distinguished from each other. The vibrational contribution (thick lines) and the sum of vibrational and electronic contributions to the free energy (thin lines) are depicted in (b), for a supercell with 54 bcc sites. The meaning of blue and red color is explained in the caption of Fig. 4. The green curve shows data obtained after applying quasi-harmonic corrections to $F_f^{vib,ZP}$. Black stars show the DFT data of Reith *et al.*[6]

Fig. 6    (Color online) Vibrational contribution (thick lines) to the free formation energy of Y (a), Ti (b), Cr (c), Mn (d), Ni (e), V (f), Mo (g), Si (h), Al (i), Co (j), and O (k) in bcc Fe. The thin lines show the sum of vibrational and electronic contributions. The color code is the same as in Figs. 4 and 5. Note that in some cases the thin and thick lines cannot be distinguished from each other. The results were obtained using a supercell with 54 bcc sites.

Fig. 7    (Color online) Vibrational contribution (thick lines) to the free formation (a) and binding (b) energy of the O-v pair calculated for a supercell containing 54 bcc sites. The sum of vibrational and electronic contributions to the free energies is depicted by thin lines which are not very different from the thick lines. The color code is the same as in Figs. 4 and 5.

Fig. 8    (Color online) Vibrational contribution (thick lines) and the sum of vibrational and electronic (thin lines) contributions to the free migration energy of the vacancy as function of temperature. The data were obtained from calculations for a supercell with 54 bcc sites. The colors are explained in preceding figure captions.

Fig.9    (Color online) Diffusivity $D(T)$ and equilibrium concentration $C_{eq}(T)$ of vacancies in bcc-Fe. The color code is the same as in previous figures. In the calculations a supercell with 54 bcc sites was considered.

Fig. 10 (Color online) Self-diffusion coefficient in bcc-Fe calculated under the assumption that the magnetization is always equal to its maximum value at $T = 0$. The meaning of the colors is explained in previous figure captions. The inset shows a table with values for $E_{SD}^{eff}$ and $D_0$ [cf. Eq. (28)], with the DFT results of Huang et al.[7] for comparison.

Fig. 11 (Color online) Experimental self-diffusion data of Hettich et al.[100], Iijima et al.[101], and Lübbehusen et al.[102] in comparison with theoretical curves obtained from the DFT data presented in Fig. 10 in combination with the semi-empirical model that considers the dependence of the spontaneous magnetization of bcc-Fe on temperature.

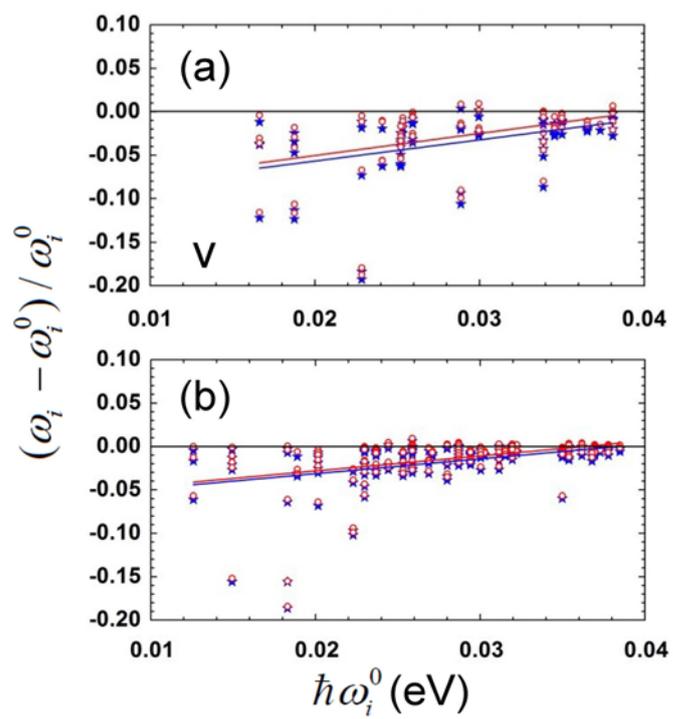

Fig. 1

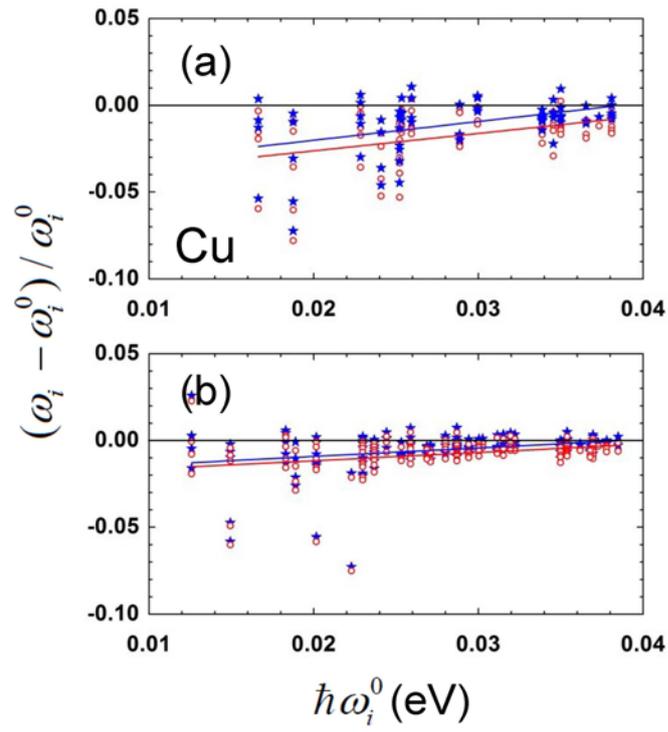

Fig. 2

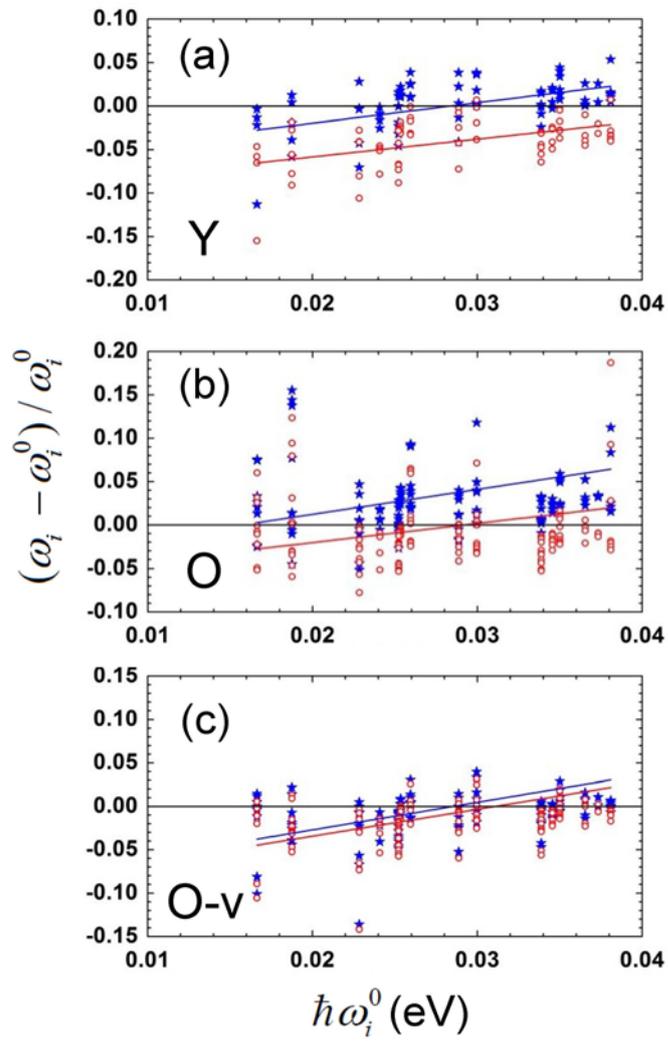

Fig. 3

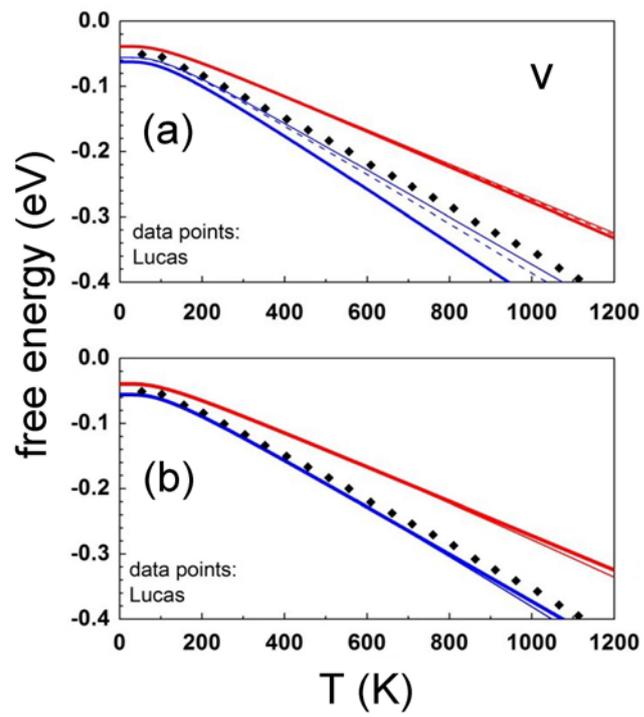

Fig. 4

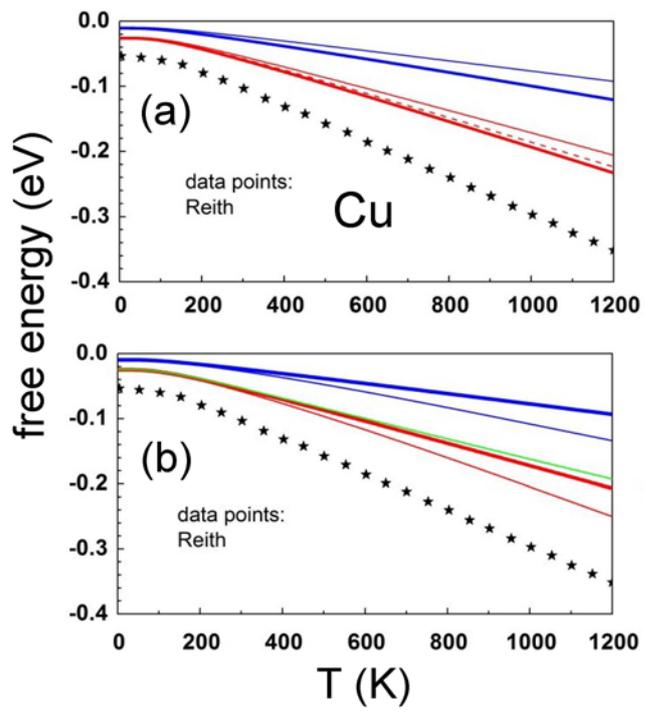

Fig. 5

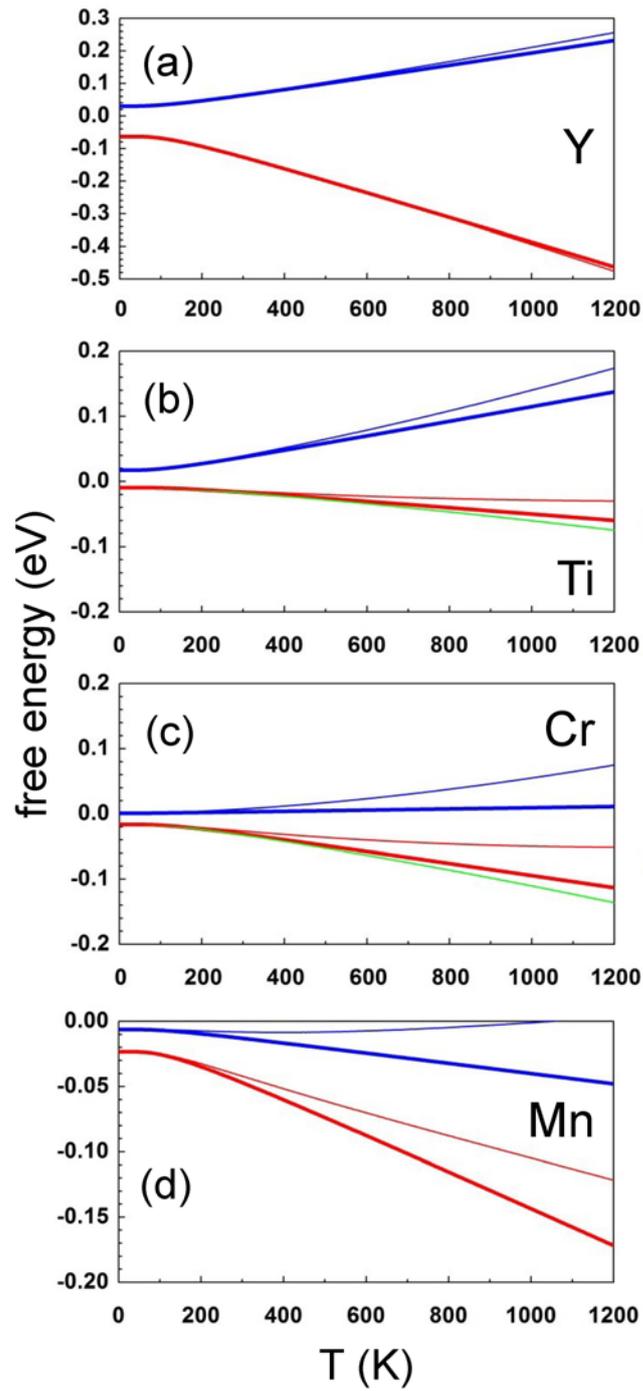

Fig. 6a-d

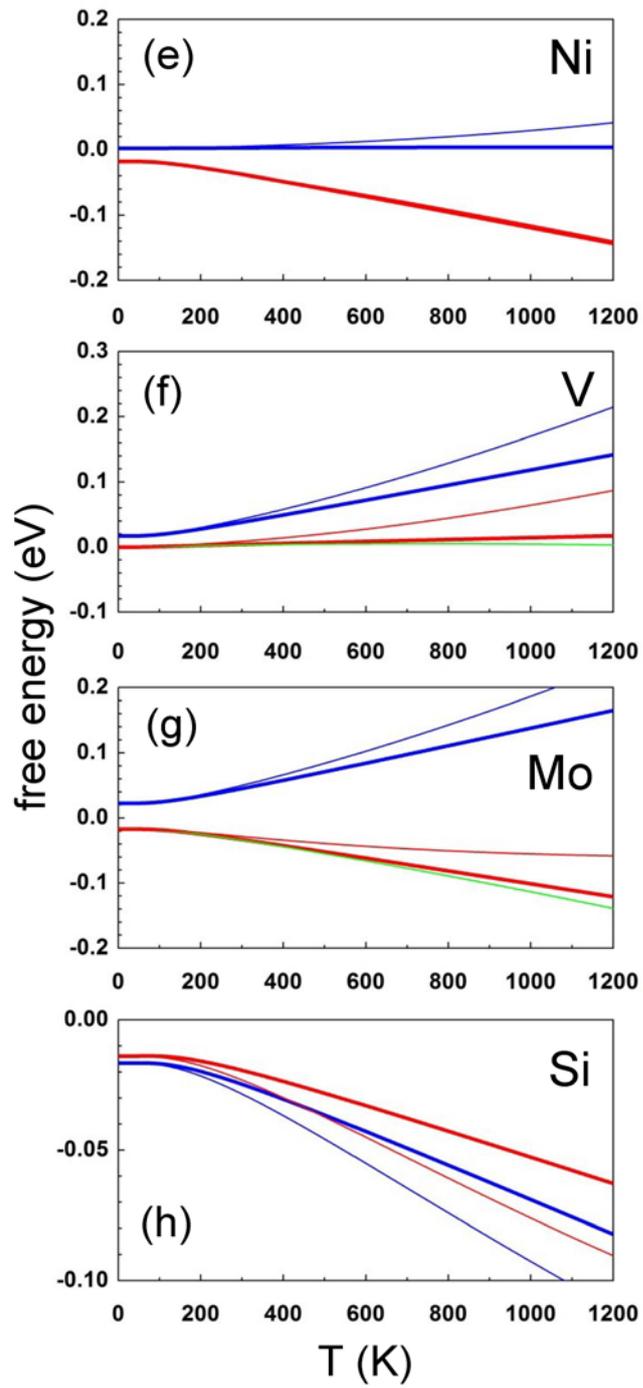

Fig. 6e-h

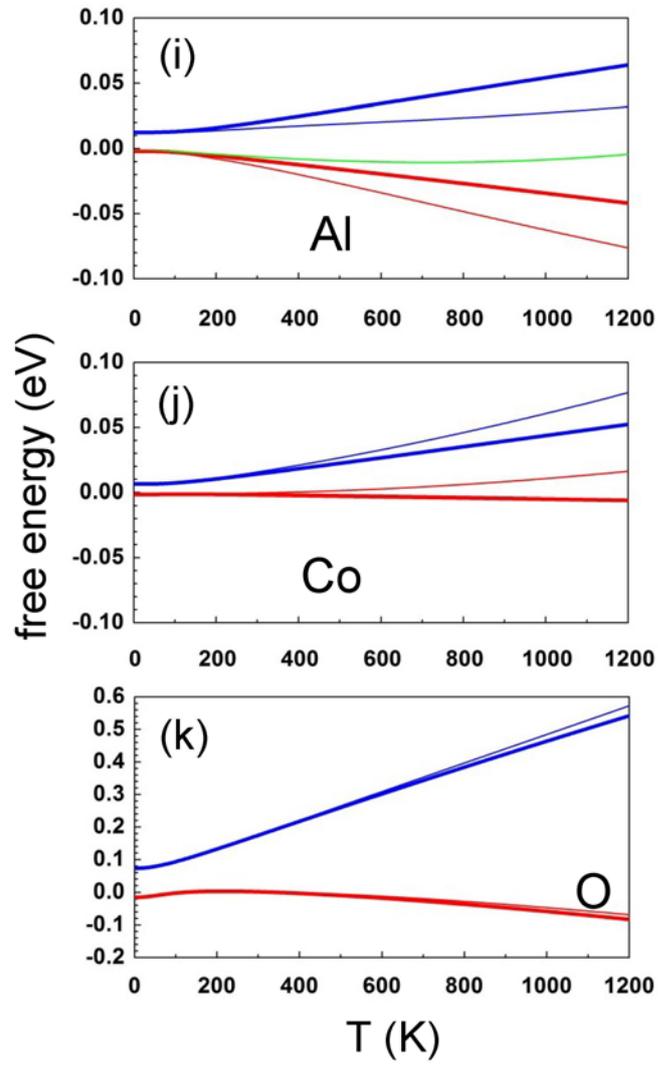

Fig. 6i-k

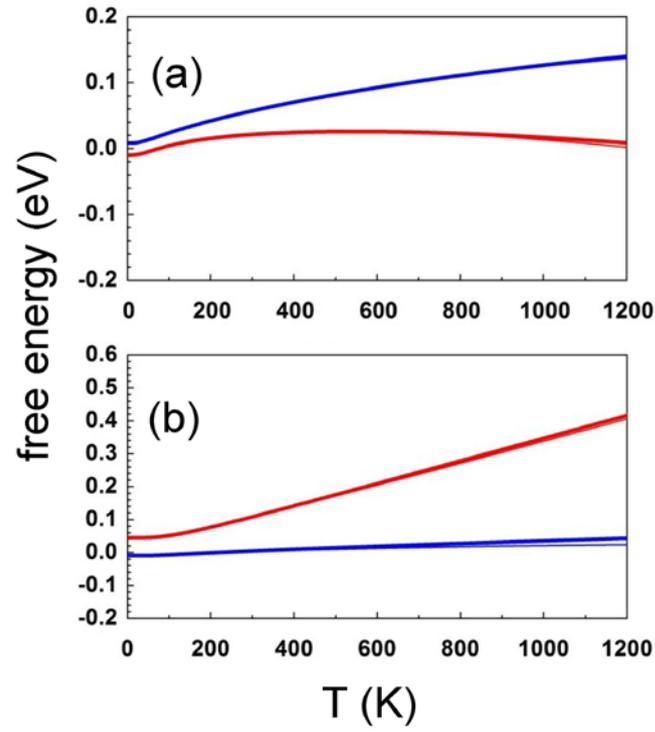

Fig. 7a-b

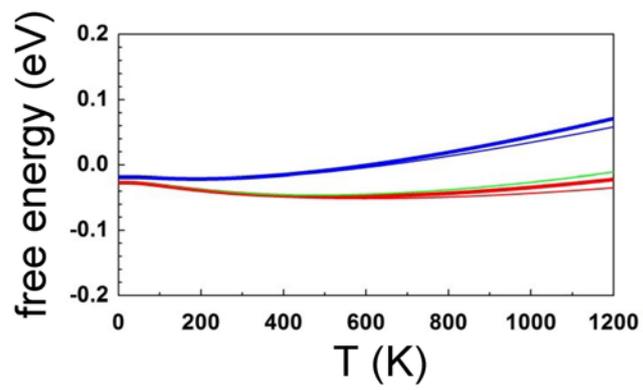

Fig. 8

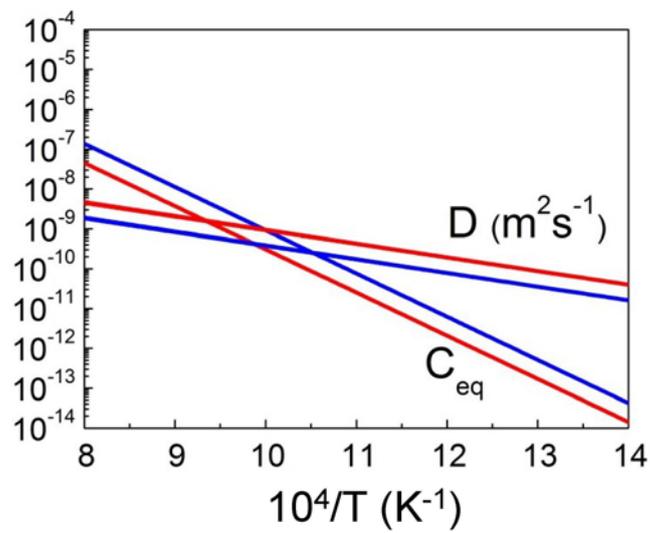

Fig. 9

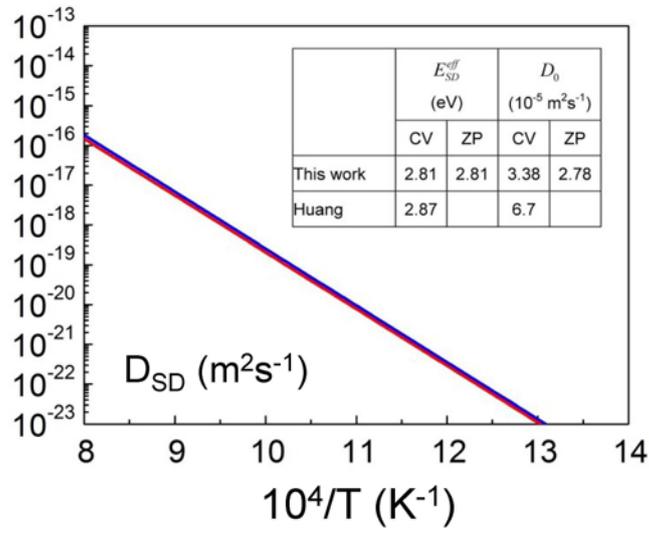

Fig. 10

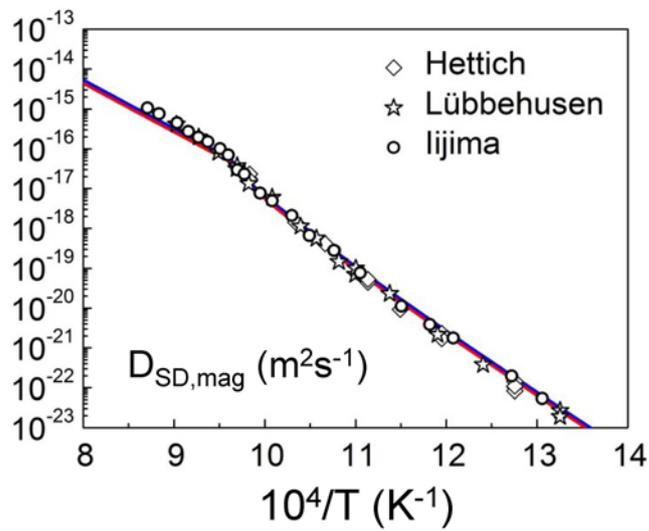

Fig. 11

**SUPPLEMENTAL MATERIAL**

**First-principles calculation of defect free energies:**

**General aspects illustrated in the case of bcc-Fe**

D. Murali, M. Posselt, and M. Schiwarth

Helmholtz-Zentrum Dresden - Rossendorf,

Institute of Ion Beam Physics and Materials Research,

Bautzner Landstraße 400, 01328 Dresden, Germany

**A. Confirmation that $E_f^{CV}$ and $E_f^{ZP}$ are fully consistent with each other [using Eq. (9)]**

The table shows $E_f^{CV} - E_f^{ZP}$ with data from Table II in comparison to the right-hand side of Eq. (9), with pressure and volume data from Table III. All data were calculated for a supercell with 54 bcc lattice sites. In the case of Si $E_f^{CV} - E_f^{ZP}$ is very small and the numerical precision of the VASP output for $p$ and $V$ is not sufficient to give a reasonable results.

| Defect | $E_f^{CV} - E_f^{ZP}$ (eV) | $\frac{V}{2} C_{ijkl} \varepsilon_{ij} \varepsilon_{kl}$ (eV) |
|---|---|---|
| v | $5.478 \times 10^{-3}$ | $6.359 \times 10^{-3}$ |
| Cu | $5.551 \times 10^{-3}$ | $5.476 \times 10^{-3}$ |
| Y | $1.987 \times 10^{-1}$ | $2.041 \times 10^{-1}$ |
| Ti | $1.575 \times 10^{-2}$ | $1.488 \times 10^{-2}$ |
| Cr | $7.435 \times 10^{-3}$ | $6.881 \times 10^{-3}$ |
| Mn | $7.601 \times 10^{-3}$ | $7.004 \times 10^{-3}$ |
| Ni | $1.135 \times 10^{-2}$ | $9.692 \times 10^{-3}$ |
| V   | $7.172 \times 10^{-3}$ | $7.049 \times 10^{-3}$ |
| Mo  | $3.714 \times 10^{-2}$ | $3.648 \times 10^{-2}$ |
| Si  | $3.270 \times 10^{-4}$ | - |
| Al  | $3.443 \times 10^{-3}$ | $4.264 \times 10^{-3}$ |
| Co  | $1.562 \times 10^{-3}$ | $1.450 \times 10^{-3}$ |
| O   | $2.123 \times 10^{-1}$ | $2.179 \times 10^{-1}$ |
| O-v | $9.493 \times 10^{-3}$ | $8.798 \times 10^{-3}$ |

**B1. Free energy of bcc-Fe, fcc-Cu, fcc-Al, and hcp-Y: Comparison with the SGTE database**

Figs. B1-1, B1-2, B1-3, and B1-4 show the comparison of the free energy data determined in this work with the SGTE database.[81,82] Note that this database was compiled using results of experimental as well as theoretical investigations and may be therefore not completely correct in regions where reliable experimental data do not exist.[81,83] Thus the aim of the comparison mainly consists in showing that the DFT data calculated in the present work are consistent with the SGTE database. The data obtained in this work and depicted in the figures were adjusted to the SGTE values at 298.15 K.

Figs. B1-1 shows the results for bcc-Fe. At first it was checked whether the restriction to the gamma point in the space of phonon wave vectors is justified. Indeed, the data of $F^{vib}(T)$ calculated for a supercell with 54 atoms are almost identical to those for a supercell with 128 and 250 atoms if the same lattice parameter is used (not shown). Furthermore, it was found that the phonon density of states determined in this work for a bcc-Fe supercell with 128 atoms is almost identical to that of Lucas *et al.*[5] who performed VASP calculations in a manner very similar to that in the present work. Fig. B1-1(a) depicts $F^{vib}(T)$ for the stress-

free supercells containing 54, 128, and 250 atoms with the lattice parameters given in Table I. Results obtained within the harmonic and the quasi-harmonic approximation [Eq. (16)] are shown. Fig. B1-1(a) illustrates that using a supercell with 54 atoms leads to sufficiently correct results. The data obtained within the quasi-harmonic approach are closer to that of the SGTE database. As discussed by other authors (cf. Ref. 28), the remaining deviations from SGTE data might be explained by anharmonic effects as well as electronic and magnetic excitations that were not considered in the calculation of the curves in Fig. B1-1(a). The curves depicted in Fig. B1-1(b) were obtained for a supercell with 54 lattice sites without and with considering the contribution of electronic excitations to the free energy. The latter effect is only visible at higher temperatures and its consideration improves the agreement with the SGTE data. The importance of the effect of electron-phonon interactions[78] at temperatures below about the half Debye temperature (200 - 250 K) was examined and no visible changes were found in any representation shown in this work.

At this point it should be noted that in Fig. B1-1 the temperature ranges up to 1200 K which is above the temperature of the ferromagnetic-to-paramagnetic transition (1043 K) and also above the temperature of the $\alpha$-to-$\gamma$-phase transition (1183 K) of iron. From the viewpoint of magnetism the presented DFT results are only valid in the ferromagnetic temperature range. However, it is important to mention that in this work the magnetization of bulk iron was always equal to its value at $T=0$, i.e. the temperature dependence of this quantity is not considered.

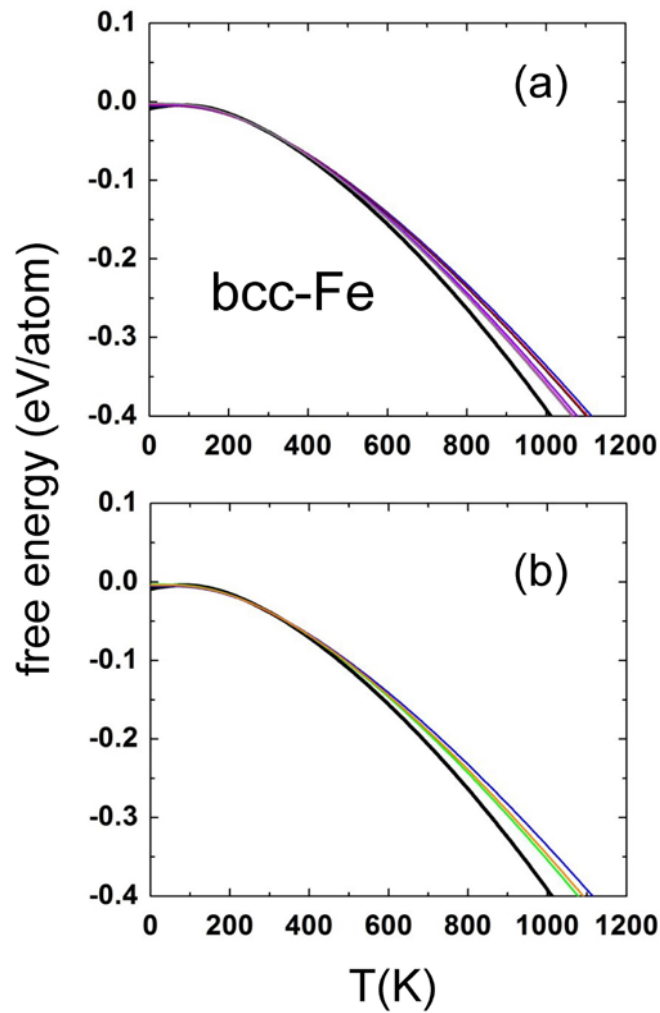

**Fig. B1-1.** Free energy per atom of bcc-Fe vs. temperature. Black lines show the data from the SGTE database.[81,82] (a): Blue and red curves depict phonon contributions to the free energy within the harmonic approximation for supercells with 54, and 128 lattice sites, respectively. Violet and magenta lines show these quantities within the quasi-harmonic approximation. In both approaches the curves for the supercell with 250 lattice sites are nearly identical to those for the supercell with 128 sites. (b): Results for a supercell with 54 bcc sites obtained within the harmonic (blue) and quasi-harmonic approach (green), and in the case where the effect of electronic excitations is taken into account additionally (orange line) within the harmonic approximation. In the quasi-harmonic approximation [Eq. (16)] the following parameters were used for bcc-Fe: $\gamma = 1.74$, $\beta = 0.00004$ K$^{-1}$, $B_0 = 170$ GPa.[87,88]

Figs. B1-2, B1-3, and B1-4 show the free energy curves for fcc-Cu, fcc-Al and hcp-Y in comparison with SGTE data.[81,82] In general a very good agreement is observed. In all cases the harmonic as well as the quasi-harmonic approach were used to determine the contributions of phonon excitations to the free energy, and the electronic contributions were taken into account additionally if phonon excitation were determined within the harmonic approximation. The importance of the effect of electron-phonon interactions[78] at temperatures below about the half Debye temperature of Cu, Al, and Y was examined and found to be negligible in the representation of Figs. B1-2, B1-3, and B1-4.

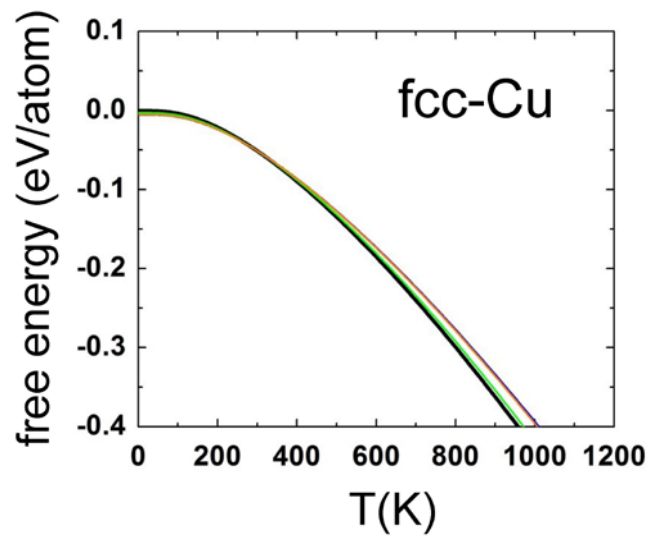

**Fig. B1-2.** Free energy per atom of fcc-Cu vs. temperature. Black lines show the data from the SGTE database.[81,82] Results for a supercell with 32 fcc sites are shown within the harmonic (blue) and quasi-harmonic approach (green) and in the case where the effect of electronic excitations was taken into account additionally (orange line) within the harmonic approximation. Note that some lines are close together. In the quasi-harmonic approximation [Eq. (16)] the following parameters were used for fcc-Cu: $\gamma = 2.0$, $\beta = 0.0000495$ K$^{-1}$, $B_0 = 138$ GPa.[88,89]

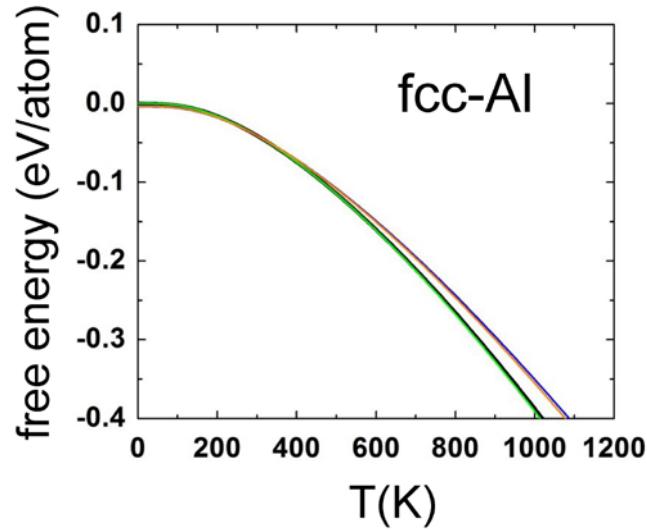

**Fig. B1-3.** Free energy per atom of fcc-Al vs. temperature. Black lines show the data from the SGTE database.[81,82] Results for a supercell with 32 fcc sites are shown. The color code is explained in the caption of Fig. B1-2. In the quasi-harmonic approximation [Eq. (16)] the following parameters were used for fcc-Al: $\gamma = 2.2$, $\beta = 0.00007$ K$^{-1}$, $B_0 = 75.9$ GPa.[88,90]

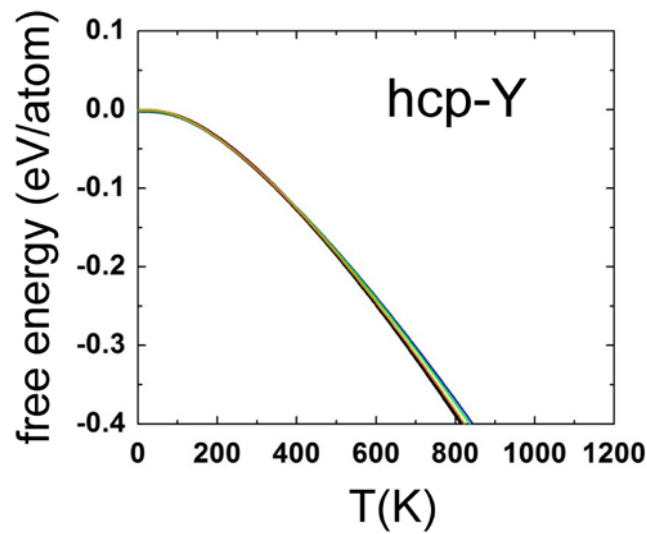

**Fig. B1-4.** Free energy per atom of hcp-Y vs. temperature. Black lines show the data from the SGTE database.[81,82] Results for a supercell with 54 hcp sites are shown. The color code is explained in the caption of Fig. B1-2. In the quasi-harmonic approximation [Eq. (16)] the following parameters were used for hcp-Y: $\gamma = 1.0$, $\beta = 0.000032$ K$^{-1}$, $B_0 = 41.2$ GPa (cf. equation on next page).[88]

In the quasi-harmonic approach, Eq. (16), the quantity $\Delta V/V$ is determined from contributions of zero-point vibrations and thermal expansion. Following e.g. Ref. 78 $\Delta V/V$ is given by

$$\frac{\Delta V}{V} = \frac{1}{B_0 V} \sum_{i=1}^{3N-3} \frac{1}{2} \hbar \omega_i^h + \beta T$$

**B2. Entropy of bcc-Fe: Comparison with recent experimental data and the SGTE database**

Fig. B2-1 shows very recent experimental data for the vibrational entropy $S^{vib}(T)$, [cf. Eq. (12)] obtained by Mauger et al.[79] using phonon densities of states measured at different temperatures. In the present work $S^{vib}(T)$ was calculated for supercells containing 54, 128 and 250 atoms within the harmonic approximation as well as within the quasi-harmonic approach using Eq. (16). The general dependence of the experimental data on temperature is well reproduced. As expected, the agreement is slightly better if the quasi-harmonic approximation and a larger supercell are used in the calculations. The remaining difference to the data points of Mauger et al.[79] is due to anharmonic contributions[79] and, possibly,[79] due to the interaction of phonons with both electronic and magnetic excitations that were not considered in present calculations, but these effects may influence the experimental data. Note that Mauger's data for $S^{vib}(T)$ are not influenced by direct electron and magnon excitations since they were obtained from the measured phonon densities of states. This should be also the main reason for the difference to the SGTE database.[81,82]

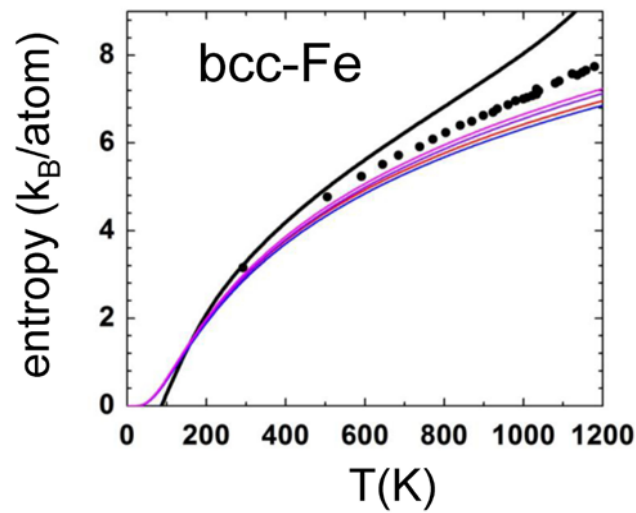

**Fig. B2-1.** Entropy per atom of bcc-Fe vs. temperature. Black lines show the data from the SGTE database.[81,82] Blue and red curves depict results obtained within the harmonic approximation for supercells with 54, 128 and 250 lattice sites, respectively. Violet and magenta lines show these quantities within the quasi-harmonic approximation. In both approaches the curves for the supercell with 250 lattice sites are nearly identical to those for the supercell with 128 sites. The data points were obtained by Mauger *et al.*[79] using phonon densities of states measured at different temperatures.

**C. Relaxation of Fe atoms in the neighbor shells of the vacancy and the Cu atom and the related force constant changes**

Fig. C-1 shows the relative change of the atomic distances with respect to the distances in perfect bcc-Fe for the vacancy and the Cu atom in a supercell with 54 bcc sites. The figures clearly show that the relative arrangement of atoms in the environment of the defects is very similar in the CV and the ZP case. Due to the additional decrease or increase of the supercell volume the relative change in the ZP case is somewhat stronger than in the CV case.

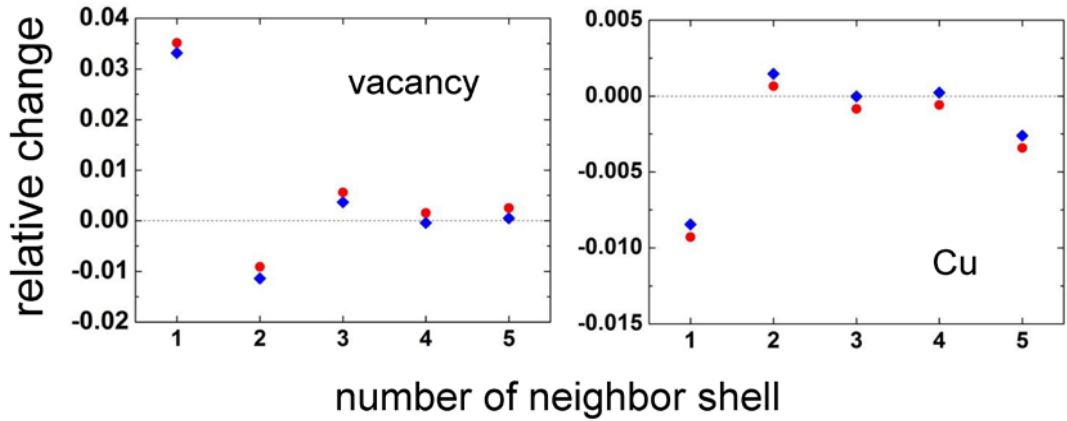

**Fig. C-1.** Relative change of atomic distances with respect to the distances in perfect bcc-Fe. Positive and negative signs denote relaxation towards and away from the defect, respectively. The blue and red symbols show results of CV and ZP, respectively. Note the different scales in the two figures.

The following table illustrates force constant data obtained from the calculated dynamical matrix. The quantities given are the forces on atom $i$ (in eV/ Å), if this atom is displaced by a given distance in the first, second, and third direction of the Cartesian coordinate system: $F_i^1 = \Phi_{ii}^{11} u_i^1$, $F_i^2 = \Phi_{ii}^{22} u_i^2$, $F_i^3 = \Phi_{ii}^{33} u_i^3$ ($u_i^1 = u_i^2 = u_i^3 = 0.015$ Å)

In the table the data for Fe atoms in the first and second neighbor shell of the vacancy and the Cu atom, and data for the Cu atom itself are given. Since the displacement is the same in all cases the data in the table are directly related to the force constants $\Phi_{ii}^{11}$, $\Phi_{ii}^{22}$, and $\Phi_{ii}^{33}$. The corresponding data for perfect bcc-Fe are given at the end of the table. Both CV and ZP data are given. The values of force constants of the Fe atoms in the first neighbor shell of both the vacancy and the Cu substitutional atom are smaller than those in bulk Fe. Furthermore, the force constants of the Cu atom are smaller than those of perfect bcc-Fe. On the average, atoms in the second neighbor shell of v and Cu have force constants close to values for perfect Fe.

Due to geometrical reasons the three values $\Phi_{ii}^{11}$, $\Phi_{ii}^{22}$, and $\Phi_{ii}^{33}$ are not equal for the atoms in the second neighbor shell.

| 1st nn Fe | displacement direction | Cu-CV | Cu-ZP | vac-CV | vac-ZP |
|---|---|---|---|---|---|
| 1 | 1 | 0.157211 | 0.154997 | 0.134803 | 0.136672 |
| 1 | 2 | 0.157203 | 0.154998 | 0.134794 | 0.136682 |
| 1 | 3 | 0.157207 | 0.154999 | 0.134693 | 0.136684 |
| 2 | 1 | 0.157202 | 0.154997 | 0.134793 | 0.136682 |
| 2 | 2 | 0.157205 | 0.154997 | 0.134795 | 0.136682 |
| 2 | 3 | 0.156974 | 0.154666 | 0.134573 | 0.136597 |
| 3 | 1 | 0.157207 | 0.155006 | 0.134807 | 0.136680 |
| 3 | 2 | 0.156978 | 0.154670 | 0.134573 | 0.136657 |
| 3 | 3 | 0.157101 | 0.154914 | 0.134722 | 0.136681 |
| 4 | 1 | 0.157202 | 0.154997 | 0.134792 | 0.136680 |
| 4 | 2 | 0.156979 | 0.154670 | 0.134571 | 0.136659 |
| 4 | 3 | 0.156975 | 0.154667 | 0.134568 | 0.136614 |
| 5 | 1 | 0.157012 | 0.154700 | 0.134554 | 0.136665 |
| 5 | 2 | 0.157205 | 0.154997 | 0.134794 | 0.136683 |
| 5 | 3 | 0.157206 | 0.154999 | 0.134692 | 0.136685 |
| 6 | 1 | 0.156980 | 0.154672 | 0.134574 | 0.136658 |
| 6 | 2 | 0.157204 | 0.154998 | 0.134792 | 0.136681 |
| 6 | 3 | 0.156975 | 0.154667 | 0.134572 | 0.136607 |
| 7 | 1 | 0.157017 | 0.154708 | 0.134556 | 0.136668 |
| 7 | 2 | 0.156976 | 0.154669 | 0.134574 | 0.136658 |
| 7 | 3 | 0.157163 | 0.154968 | 0.134724 | 0.136677 |
| 8 | 1 | 0.156977 | 0.154672 | 0.134574 | 0.136654 |
| 8 | 2 | 0.156976 | 0.154669 | 0.134575 | 0.136655 |
| 8 | 3 | 0.156976 | 0.154669 | 0.134572 | 0.136613 |
| Cu atom | | | | | |
| | 1 | 0.132428 | 0.130578 | - | - |
| | 2 | 0.132428 | 0.130604 | - | - |
| | 3 | 0.132423 | 0.130600 | - | - |
| 2nd nn Fe | | | | | |
| 1 | 1 | 0.149558 | 0.147806 | 0.151603 | 0.154170 |
| 1 | 2 | 0.166042 | 0.163646 | 0.167762 | 0.170362 |
| 1 | 3 | 0.166325 | 0.163932 | 0.168016 | 0.170494 |
| 2 | 1 | 0.166330 | 0.163935 | 0.168021 | 0.170487 |
| 2 | 2 | 0.149558 | 0.147806 | 0.151708 | 0.154163 |
| 2 | 3 | 0.166044 | 0.163647 | 0.167771 | 0.170362 |
| 3 | 1 | 0.166335 | 0.163942 | 0.168019 | 0.170499 |
| 3 | 2 | 0.166524 | 0.164133 | 0.168118 | 0.170500 |
| 3 | 3 | 0.149964 | 0.148218 | 0.151764 | 0.154176 |
| 4 | 1 | 0.166094 | 0.163943 | 0.167829 | 0.170357 |
| 4 | 2 | 0.166317 | 0.163920 | 0.168017 | 0.170571 |
| 4 | 3 | 0.150386 | 0.148598 | 0.153001 | 0.154265 |
| 5 | 1 | 0.166513 | 0.164118 | 0.167996 | 0.170504 |
| 5 | 2 | 0.150389 | 0.148602 | 0.152999 | 0.154314 |
| 5 | 3 | 0.166054 | 0.163656 | 0.167738 | 0.170398 |
| 6 | 1 | 0.150417 | 0.148626 | 0.153024 | 0.154188 |
| 6 | 2 | 0.166057 | 0.163659 | 0.167741 | 0.170399 |
| 6 | 3 | 0.166324 | 0.163929 | 0.168013 | 0.170497 |
| | | | | | |
| Bulk Fe | | | | | |
| | 1 | 0.165117 | - | - | - |
| | 2 | 0.165298 | - | - | - |
| | 3 | 0.165309 | - | - | - |

The table below shows the relative change of average force constants with respect to the value in perfect bcc-Fe, for the first and second neighbor shells. The relative change resulting from the replacement of a Fe atom by a Cu atom is also shown. A negative and positive sign denotes a relative decrease and increase, respectively. Since for atoms in the first neighbor shell the data obtained from displacements in the three different directions are nearly equal (see table above) the averaging was performed over all atoms and directions. For the second neighbor atoms the averaging was performed over the atomic data with the smaller and the two larger absolute values separately.

|  | **Cu-CV** | **Cu-ZP** | **vac-CV** | **vac-ZP** |
|---|---|---|---|---|
| **1st nn Fe** |  |  |  |  |
| 1 | -0.0485 | -0.0618 | -0.1836 | -0.1697 |
|  |  |  |  |  |
| **2nd nn Fe** |  |  |  |  |
| 1 | -0.0909 | -0.1030 | -0.0787 | -0.0666 |
| 2 | 0.0061 | -0.0061 | 0.0181 | 0.0303 |
|  |  |  |  |  |
| **Cu atom** |  |  |  |  |
| 1 | -0.2000 | -0.2121 | - | - |

**D. Values of the parameter $\gamma'$ determined from Eq. (17) using the volume data given in Table III**

| Defect | $\gamma'$ |
|---|---|
| | **54 sites** |
| v | 1.7252 |
| Cu | 1.7380 |
| | **128 sites** |
| v | 1.7399 |
| Cu | 1.7043 |
| | **250 sites** |
| v | 1.7579 |
| Cu | 1.7259 |
| | **54 sites** |
| Y | 1.7420 |
| Ti | 1.8614 |
| Cr | 1.7154 |
| Mn | 1.6697 |
| Ni | 1.6968 |
| V | 1.7216 |
| Mo | 1.7324 |
| Si | 1.8818 |
| Al | 1.7653 |
| Co | 1.7371 |
| O | 1.6418 |
| O-v | 1.6912 |

An average over all data gives the value $\gamma' = 1.74$.

## E. Material parameters used in the calculation of defect free energies within the quasi-harmonic approach

The phonon Grüneisen parameter $\gamma$ is determined using the values of the volume expansion coefficient $\beta$, the bulk modulus $B_0$, the density $\rho$, and the specific heat capacity $c_s$ given in Ref. 88, with

$$\gamma = \frac{\beta B_0}{\rho c_s}$$

| Material | $\beta$ ($10^{-5}$ K$^{-1}$) | $B_0$ (GPa) | $\rho$ (g cm$^{-3}$) | $c_s$ (J g$^{-1}$K$^{-1}$) | $\gamma$ |
|---|---|---|---|---|---|
| Al | 6.93 | 75.9 | 2.70 | 0.897 | 2.2 |
| Cu | 4.95 | 138 | 8.93 | 0.385 | 2.0 |
| Cr | 1.47 | 152 | 7.15 | 0.449 | 0.7 |
| Ni | 4.02 | 186 | 8.90 | 0.444 | 1.9 |
| V | 2.52 | 155 | 6.00 | 0.489 | 1.3 |
| Mo | 1.44 | 260 | 10.2 | 0.251 | 1.45 |
| Co | 3.90 | 191 | 8.86 | 0.421 | 2.0 |
| Ti | 2.58 | 105 | 4.51 | 0.523 | 1.15 |
| Y | 3.20 | 41.2 | 4.47 | 0.30 | 1.0 |

## F. Confirmation that $F_f^{\text{vib,CV}}$ and $F_f^{\text{vib,ZP}}$ are fully consistent with each other

2$^{\text{nd}}$ and 3$^{\text{rd}}$ column: Difference between CV- and ZP-based vibrational contributions to the free formation energies at $T = 0$ obtained from Figs. 4-7 compared to the corresponding values determined by Eq. (18), using data for $\gamma'$ given in Sec. D of this Supplement and volume data from Table III.

6$^{\text{th}}$ and 7$^{\text{th}}$ column: Difference between CV- and ZP-based vibrational contributions to formation entropies from Figs. 4-7 (for $T > 500$ K) in comparison with data calculated by Eq. (21) using volume data from Table III as well as high-temperature values for $\beta$ and $B_0$ of bcc-Fe ($\beta = 0.00004$ K$^{-1}$, $B_0 = 155$ GPa, cf. Ref. 87). All data were calculated for a supercell with 54 bcc lattice sites.

| Defect | $F_f^{\text{vib,ZP}}(0) - F_f^{\text{vib,CV}}(0)$ (eV) | $-\gamma' \dfrac{V-V_0}{V_0} \sum_{i=1}^{3N-3} \dfrac{1}{2} \hbar \omega_i^{CV}$ (eV) | $S_f^{\text{vib,ZP}} - S_f^{\text{vib,CV}}$ ($k_B$) | $\beta B_0 (V - V_0)$ ($k_B$) |
|---|---|---|---|---|
| v | $1.630 \times 10^{-2}$ | $1.612 \times 10^{-2}$ | -1.143 | -1.199 |
| Cu | $-1.545 \times 10^{-2}$ | $-1.520 \times 10^{-2}$ | 1.075 | 1.084 |
| Y | $-9.327 \times 10^{-2}$ | $-9.271 \times 10^{-2}$ | 6.580 | 6.515 |
| Ti | $-2.70 \times 10^{-2}$ | $-2.695 \times 10^{-2}$ | 1.871 | 1.769 |
| Cr | $-1.716 \times 10^{-2}$ | $-1.686 \times 10^{-2}$ | 1.175 | 1.204 |
| Mn | $-1.71 \times 10^{-2}$ | $-1.679 \times 10^{-2}$ | 1.172 | 1.238 |
| Ni | $-2.03 \times 10^{-2}$ | $-1.993 \times 10^{-2}$ | 1.376 | 1.444 |
| V | $-1.67 \times 10^{-2}$ | $-1.677 \times 10^{-2}$ | 1.180 | 1.192 |
| Mo | $-3.95 \times 10^{-2}$ | $-3.880 \times 10^{-2}$ | 2.707 | 2.732 |
| Si | $2.67 \times 10^{-3}$ | $2.771 \times 10^{-3}$ | -0.1839 | -0.1897 |
| Al | $-1.45 \times 10^{-2}$ | $-1.428 \times 10^{-2}$ | 1.003 | 0.9894 |
| Co | $-8.04 \times 10^{-3}$ | $-7.812 \times 10^{-3}$ | 0.5516 | 0.5517 |
| O | $-8.849 \times 10^{-2}$ | $-8.528 \times 10^{-2}$ | 6.155 | 6.100 |
| O-v | $-1.029 \times 10^{-2}$ | $-1.769 \times 10^{-2}$ | 1.401 | 1.280 |